%% file: paper_revised.tex
\documentclass[twocolumn,aps,prx,longbibliography,superscriptaddress]{revtex4-2}
\usepackage{amsmath}
\usepackage{graphicx}
\usepackage{hyperref}
\usepackage[normalem]{ulem}
\usepackage{bbold}
\hypersetup{colorlinks=true, linkcolor=blue, citecolor=blue, urlcolor=blue}

\begin{document}
\title{Sublinear scaling in non-Markovian open quantum systems simulations}
\author{Moritz Cygorek}
\affiliation{SUPA, Institute of Photonics and Quantum Sciences, Heriot-Watt University, Edinburgh EH14 4AS, United Kingdom}
\author{Jonathan Keeling}
\affiliation{SUPA, School of Physics and Astronomy, University of St Andrews, St Andrews KY16 9SS, United Kingdom}
\author{Brendon W. Lovett}
\affiliation{SUPA, School of Physics and Astronomy, University of St Andrews, St Andrews KY16 9SS, United Kingdom}
\author{Erik M. Gauger}
\affiliation{SUPA, Institute of Photonics and Quantum Sciences, Heriot-Watt University, Edinburgh EH14 4AS, United Kingdom}

\begin{abstract}
While several numerical techniques are available for predicting the dynamics of 
non-Markovian open quantum systems,
most struggle with simulations 
for very long memory and propagation times, e.g., 
due to superlinear scaling with the number of time steps $n$. 
Here, we introduce a numerically exact algorithm to calculate 
process tensors---compact representations of environmental influences---which 
provides a scaling advantage
over previous algorithms by leveraging self-similarity of the tensor networks
that represent the environment. It is applicable to environments with 
Gaussian statistics, such as for spin-boson-type open quantum systems.
Based on a divide-and-conquer strategy, our approach requires only 
$\mathcal{O}(n\log n)$  singular value decompositions for 
environments with infinite memory.  Where the memory can be truncated after $n_c$ time steps, a nominal scaling $\mathcal{O}(n_c\log n_c)$ is found, which is independent of $n$.  This improved scaling is enabled  by identifying 
process tensors with repeatable blocks.
To demonstrate the power and utility of our approach we provide three examples. 
(1) We calculate the fluorescence  spectra of a quantum dot
under both strong driving and
strong dot-phonon couplings, a task requiring simulations over millions of
time steps, which we are able to perform in minutes. 
(2) We efficiently find
process tensors describing superradiance of multiple emitters.
(3) We explore the limits of our algorithm by considering coherence decay
with a very strongly coupled environment. 
The observed computation time is not necessarily proportional
to the number of singular value decompositions, because the 
matrix dimensions also depend on number of time steps. Nevertheless, 
quasi-linear and sublinear scaling of computation time is found in practice 
for a wide range of parameters.
The algorithm we present here not only significantly extends the scope of 
numerically exact techniques to open quantum systems with long 
memory times, but also has fundamental implications for the simulation
complexity of tensor network approaches.
\end{abstract}

\maketitle

\section{Introduction}
A common challenge in quantum technology is the ubiquity of 
dephasing and dissipation caused by interactions between quantum systems
and their surrounding environment~\cite{BreuerPetruccione}. 
Understanding environmental influences is, thus, crucial
for mitigating them~\cite{PT_White2020,Reiter_distinctive_characteristics,
Reiter_review2019, Kaldewey2017,NARP_Hall}, 
or even using them to enhance the 
functionality of quantum devices,  as in phonon-assisted state 
preparation~\cite{QuilterPRL, PI_singlephoton} or environment-assisted 
quantum transport~\cite{Maier_ENAQT,Chin_noise-assisted,Plenio_dephasing-assisted}.
A standard approach for modeling the dynamics of open quantum systems is the
Lindblad master equation~\cite{BreuerPetruccione,Lindblad}. This can be
derived using perturbation theory and the Born-Markov approximation, 
which assumes that the environment is static and memoryless and couples only weakly to the system. 
However, this is often an oversimplification
and more sophisticated methods are required to accurately model the dynamics
in many cases, where the structure of the environment matters or coupling is strong. Examples include: solid-state quantum dots (QDs) coupled to local phonon 
environments~\cite{Reiter_review2019,CoopWiercinski,Denning_polaron_polariton,
Hughes_polaron2012, Review_Nazir}, 
charge or excitation transport in biomolecules~\cite{Chin_noise-assisted,
Makri_ring,T-TEDOPA}, 
ultrafast spin dynamics~\cite{DMS_trend_reversal,DMS_nonmagnetic},
spontaneous emission in photonic 
structures~\cite{Hoeppe_photonic,John_photonic},
and superconducting qubits in quantum computers~\cite{PT_White2020}.

Beyond the restrictive weak-coupling Born-Markov approximation, 
the regime of non-Markovian dynamics~\cite{RevModPhys_deVega} arises. 
Here a quantum system triggers a dynamical evolution of the environment, and
the environment's response feeds back to the system 
at a later point in time, constituting a time-non-local memory.
Several strategies have been discussed for modeling non-Markovian 
dynamics~\cite{RevModPhys_deVega}.
One general strategy is to solve the Schr{\"o}dinger equation
for the total closed system composed of the system of interest 
and its environment.
Because environments typically consist of a large number---often
a continuum---of degrees of freedom, this becomes numerically intractable
unless the model is somehow reduced to a finite number of effective degrees of
freedom. How to choose the relevant degrees of freedom 
differs from method to method:
Mean-field, cumulant, or cluster expansions~\cite{RevModPhys_Kuhn,Axt_DCT} 
are based on the assumption that higher-order correlations are negligible,
and thus one may  consider only degrees of freedom within the subspace of product 
states, or states with low-order correlations.
Other methods based on techniques from many-body quantum theory 
employ efficient representations of the total system 
such as multilayer multiconfiguration time-dependent Hartree 
wave functions~\cite{MLMCTDH_anharmonic} 
or matrix product states~\cite{T-TEDOPA,DAMPF,PRX_Goold}. 
In the reaction coordinate mapping~\cite{Nazir_RC_NJP2022}, parts
of the environment are explicitly taken into account by extending the system
of interest. A similar concept is to replace the actual environment
by a small set of auxiliary oscillators~\cite{AuxiliaryOscillators}.

A second general strategy for simulating open quantum systems  is to 
keep the description confined to the degrees of freedom of the 
system, 
but to account for environmental effects via equations of motion 
that are non-local in time. 
For example, the Nakajima-Zwanzig formalism~\cite{Nakajima,BreuerPetruccione} 
explicitly includes memory effects of the form of an integral over past times.
The Feynman-Vernon path integral formalism~\cite{FeynmanVernon} 
encapsulates environmental effects in an influence functional, which
acts as a trajectory-dependent weight in a sum over all possible system 
trajectories.
This formalism is the starting point for several practical methods for open quantum systems simulation.
For Gaussian environments with exponentially decaying bath correlations, repeated time-differentiation of the Feynman-Vernon path sum gives rise to a set of Hierarchical Equations of Motion (HEOM)~\cite{HEOM89,HEOM90}. Alternatively, the path sum can be expressed as a stochastic average with bath correlations encoded in the noise statistics~\cite{Strunz_diffusion}, from which one can derive non-Markovian stochastic Schr{\"o}dinger equations~\cite{Diosi_stochastic,Diosi_diffusion,stochastic_Stockburger}. The combination of both ideas leads to equations for a Hierarchy of Pure States (HOPS)~\cite{Suess_Hierarchy}. These concepts can be further generalized, e.g., to stochastic master equations for non-Gaussian environments~\cite{stochastic_HsiehI} and to open quantum systems simulations where the environment is continuously measured~\cite{Daley_cHEOM}.

Here, we focus on the question of how to predict non-Markovian dynamics
for both long propagation times and long memory times.
While some specific features of long time dynamics---such as the nature of the steady state---may sometimes be more accessible straightforwardly, the problem of modeling non-Markovian \emph{dynamics} over long times is important but challenging. For example, one may want to find the general time evolution, or multi-time correlations of a system, but 
in all the methods described above, it is generally hard to track 
such non-Markovian dynamics over long times.
Even though strategies of explicitly representing environment degrees of
freedom at first glance seem to scale linearly
with the total number of time steps $n$, i.e.\  $\mathcal{O}(n)$, they tend to
become inaccurate with increasing propagation time. 
This is due to the increase with time of the number
of degrees of freedom of the environment that can be excited and thus entangled with the system. 
For example, consider the discretization of a continuum
of phonon modes by energy intervals of width $\hbar\Delta \omega$. 
Propagating over a time $t_e$, energy-time uncertainty implies
that a discretization of $\Delta\omega \lesssim 1/t_e$ is required 
to preclude resolving a difference between the continuum and the
discretized model. 
The need for a finer discretization results in a scaling of 
$\mathcal{O}(n^2)$ unless a more sophisticated, e.g., adaptive, discretization
is used, or the resolution of individual sampling points is blurred by 
some form of line broadening~\cite{DAMPF}. 
In approaches where the environment is mapped to a chain of coupled oscillators~\cite{T-TEDOPA,DAMPF}, 
this statement is related to the observation that the Lieb-Robinson theorem restricts
environment excitation to a light cone~\cite{Plenio_LiebRobinson},
which increases in size with increasing propagation time $t_e$.
Similarly, stochastic sampling requires more trajectories to reach the 
same accuracy when the propagation time $t_e$ is increased. 
Hence, most numerical methods for open quantum systems show superlinear
scaling with the number of time steps $n$.

For direct applications of the Feynman-Vernon influence functional approach,
one finds exponential scaling
with $n$. However, a reformulation into the iterative scheme QUAPI~\cite{QUAPI1, QUAPI2}
yields truly linear scaling when the memory time of the environment is finite
and so the memory can be truncated after $n_c$ time steps. 
Such an approach however retains exponential scaling with the memory time $n_c$.
It could be expected that $\mathcal{O}(n)$ is the natural lower bound for the
scaling, in particular for general open quantum systems 
with time dependent driving, because, at the very least, $\mathcal{O}(n)$
observables have to be obtained in order to extract the full time evolution. 
In practice, however, it may happen that the numerically most demanding 
parts of the simulation can be precalculated in sublinear time, leaving
a linear-in-time scaling with very small prefactor 
for the evaluation of the full time evolution which is irrelevant
for most practical applications. 
An example is the small matrix decomposition of 
the path integral (SMatPI) approach~\cite{SMatPI},
where, for a fixed time-independent Hamiltonian, an effective propagator
in Liouville space including the effects of the environment is obtained,
which can be used to propagate a system over time at a numerical cost 
comparable to a Lindblad master equation. 
Yet, for applications involving time-dependent Hamiltonians, e.g., 
to simulate the response of open quantum systems to strong driving with 
shaped laser pulses~\cite{pulsed_spectra,SUPER_entanglement}, the effective
propagator would have to be recalculated for each time step.

In this paper, we introduce an algorithm based on the process tensor (PT)
framework that demonstrates sublinear scaling of the numerically demanding step. 
A PT captures environment influences equivalent to the the Feynman-Vernon influence functional in a numerically exact way and can be efficiently represented in the form of a matrix product operator (MPO)~\cite{PT_PRA,JP}.
Once obtained, an open quantum system with an arbitrary time-dependent system
Hamiltonian can be propagated in time by simple
matrix multiplications on a vector space given by the product of the system Liouville space and the inner dimension of the PT-MPO, which we denote as $\chi$.
The bottleneck in MPO techniques is the MPO compression, i.e., the reduction
of the inner dimension $\chi$, using rank-reducing operations, often achieved by
truncated singular value decomposition (SVD).
We show that the self-similarity of time-independent Gaussian environments---which include
electromagnetic environments, typical vibrational baths, 
and the paradigmatic spin-boson model---can be exploited to devise 
a divide-and-conquer scheme which 
reduces the number of SVDs from $\mathcal{O}(n^2)$ in the 
algorithm introduced by J{\o}rgensen and Pollock~\cite{JP} to $\mathcal{O}(n\log n)$. 
If the memory time is further limited to $n_c$ time steps, we arrive at
a theoretical scaling $\mathcal{O}(n_c\log n_c)$, constant in $n$.
The actual scaling can be larger than this, as the bond dimension $\chi$, and thus the time required for SVD evaluation, can depend on the propagation time.  However, we will see from numerical results that sublinear scaling with $n$ is indeed seen in the examples we study.

To test the performance of our algorithm in practice and to demonstrate
a sample of new applications it enables, we provide several examples: 
First, we investigate the fluorescence spectra of semiconductor QDs coupled
to a bath of acoustic phonons with strong driving. While being of considerable interest 
for experiments~\cite{Michler_Mollow2011,Michler_Mollow2012}, 
numerically calculating QD fluorescence spectra is a challenging multi-scale problem,
because the width of the zero-phonon line is determined by radiative lifetimes
of the order of nanoseconds, while typical phonon memory times are of the
order of picoseconds, and strong driving forces us to use small time steps
on the femtosecond scale. The sublinear scaling of our algorithm enables
us to obtain numerically exact spectra from simulations involving 
a million time steps within minutes on a conventional laptop computer. 
Second, we use the capability of our algorithm to deal with very many
time steps to study superradiance of multiple emitters without making any rotating wave 
approximation. This allows us to describe the breakdown of superradiance in the 
presence of disorder and dephasing due to interactions with phonon environments.
Finally, we discuss  coherence decay for a system  with 
strong coupling to an environment with a strongly  peaked spectral density. This example
illuminates where the limitations of our algorithm arise.

The article is structured as follows:
In the theory section~\ref{sec:theory}, 
we introduce and describe our algorithm, where, in subsection~\ref{sec:PT},
we first summarize the PT formalism, on which our algorithm is based. 
For comparison with the commonly-used sequential algorithm by 
J{\o}rgensen and Pollock~\cite{JP},
and to introduce quantities also used in our approach,
we revise the PT calculation scheme for Gaussian environments in Ref.~\cite{JP}
in subsection~\ref{sec:JP}. Then, we introduce our divide-and-conquer scheme
in subsection~\ref{sec:DnC} and introduce
periodic PTs in subsection~\ref{sec:periodicPT}.
The examples of fluorescence spectra, multi-emitter superradiance, and
coherence decay are discussed in 
subsections~\ref{sec:QDPhonon}, \ref{sec:superrad}, and \ref{sec:coherence_decay}, 
respectively. 
Our results are summarized in section~\ref{sec:summary}.

\section{\label{sec:theory}Theory}
\subsection{\label{sec:PT}Process Tensors}
An open quantum system is  defined by dividing the total system into 
a system of interest $S$ and an environment $E$. Correspondingly,
the total Hamiltonian is decomposed into  $H=H_S+H_E$, where $H_S$
is an arbitrary, possibly time-dependent, system Hamiltonian and 
$H_E$ is the environment Hamiltonian, which also includes the interaction
with the system of interest.
In this paper, we present an algorithm applicable to environments
with Gaussian correlations, focusing on a generalized spin-boson model
consisting of a few-level system 
coupled linearly to a bath of harmonic oscillators, described by 
\begin{align}
\label{eq:spinboson}
H_E=\sum_k \hbar\omega_k b^\dagger_k b_k 
+\sum_k \hbar (g_k b^\dagger_k + g^*_k b_k )\hat{O},
\end{align} 
where $\hat{O}$ is a Hermitian operator acting on the
system Hilbert space. 
The environment Hamiltonian is conveniently characterized
by the spectral density $J(\omega)=\sum_k |g_k|^2 \delta(\omega-\omega_k)$. 
Here, we choose to work in the basis where the coupling 
is diagonal $\hat{O}=\sum_u \lambda_u |u\rangle\langle u|$ with a set of 
basis states $|u\rangle$ of the system Hilbert space.

Throughout this article, we use a compact Liouville space notation
and assume a uniform time grid $t_j=t_0+ j\Delta t$ with time step $\Delta t$.
For example, the reduced system density matrix 
${\bar{\rho}}(t)=\textrm{Tr}_E[{\rho}(t)]$
at time $t_j$ is denoted by 
$\bar{\rho}_{\alpha_j}=\bar{\rho}_{s_j,r_j}(t_j)=
\langle s_j| \bar{\rho}(t)|r_j\rangle$,
where $s_j$ and $r_j$ are system Hilbert space indices, which are
combined into the Liouville space index $\alpha_j=(s_j,r_j)$.
The time argument $t_j$ is implied in the index $\alpha_j$.
In this notation, the reduced density matrix $\bar{\rho}_{\alpha_n}$ at
a final time $t_n$ can be expressed as~\cite{FeynmanVernon}:
\begin{align}
\bar{\rho}_{\alpha_n}=& \sum_{\alpha_{n-1},\dots,\alpha_0} 
\mathcal{F}^{\alpha_n,\dots,\alpha_1}
\bigg(\prod_{l=1}^{n} \mathcal{M}_{\alpha_l, \alpha_{l-1}}\bigg)
\bar{\rho}_{\alpha_0},
\label{eq:feynmanvernon}
\end{align}
where $\mathcal{M}_{\alpha_l, \alpha_{l-1}}= 
\big( e^{\mathcal{L}_S\Delta t} \big)_{\alpha_l, \alpha_{l-1}}$ describes the
free evolution of the system during one time step $\Delta t$ 
under the free system Liouvillian, e.g., 
$\mathcal{L}_S[\bar{\rho}]=\frac i\hbar \big[H_S, \bar{\rho}\big]$ and
$\mathcal{F}^{\alpha_n,\dots,\alpha_1}$ is the 
Feynman-Vernon influence functional~\cite{FeynmanVernon}, which exactly
captures all environment influences in the limit $\Delta t\to0$.

For the generalized spin-boson model, the influence functional is given by~\cite{TEMPO,JP}
\begin{align}
\mathcal{F}^{\alpha_n,\dots,\alpha_1}=&
\prod_{i=1}^n \prod_{j=1}^i \big[b_{i-j}\big]^{\alpha_i,\alpha_j}
=\prod_{j=1}^n \prod_{l=0}^{n-j} \big[b_{l}\big]^{\alpha_{l+j},\alpha_{j}},
\label{eq:IF}
\end{align}
where the factors $\big[b_{i-j}\big]^{\alpha_i,\alpha_j}$ are related to 
the bath correlation function via
\begin{align}
\big[b_{(i-j)}\big]^{\alpha_i,\alpha_j}=& 
e^{-(\lambda_{s_i}-\lambda_{r_i})
(\eta_{i-j}\lambda_{s_j} -\eta^*_{i-j}\lambda_{r_j})}
\label{eq:b}
\end{align}
with $\alpha_j=(s_j,r_j)$ and $\lambda_{s,r}$ as defined above, and~\footnote{Note that the time integrals in Eq.~\eqref{eq:eta} 
can be solved analytically and the remaining integral over $\omega$ 
in Eq.~\eqref{eq:BCF} can be expressed as a Fourier transform, 
so that $n$ values of $\eta_{i-j}$ are efficiently 
obtained in $\mathcal{O}(n\log n)$ time.}
\begin{align}
\label{eq:eta}
\eta_{i-j}=& \begin{cases} \int_{t_{i-1}}^{t_i} dt' 
\int_{t_{i-1}}^{t'}dt''\, C(t'-t''), & i=j \\
\int_{t_{i-1}}^{t_i}dt' 
\int_{t_{j-1}}^{t_j}dt''\, C(t'-t''), & i\neq j 
\end{cases}
\end{align}
where
\begin{align}
C(t)=&\int_0^\infty d\omega\, J(\omega)\Big[
\textrm{coth}\big(\tfrac 12\beta\hbar\omega\big) \cos(\omega t) 
- i\sin(\omega t)\Big].
\label{eq:BCF}
\end{align}
The memory of the environment is finite if the bath correlation function
$C(t)$ vanishes after $n_c$ time steps, i.e.~\mbox{$C(t\ge n_c\Delta t)\approx 0$,} 
which implies $\eta_{l \ge n_c}\approx 0$ and  
$\big[b_{l\ge n_c}\big]^{\alpha_i,\alpha_j}\approx 1$. 
In this case, it is sufficient to consider at most $n_c$ terms 
in the second product in the right hand side of Eq.~\eqref{eq:IF}.

\begin{figure*}
\includegraphics[width=0.98\linewidth]{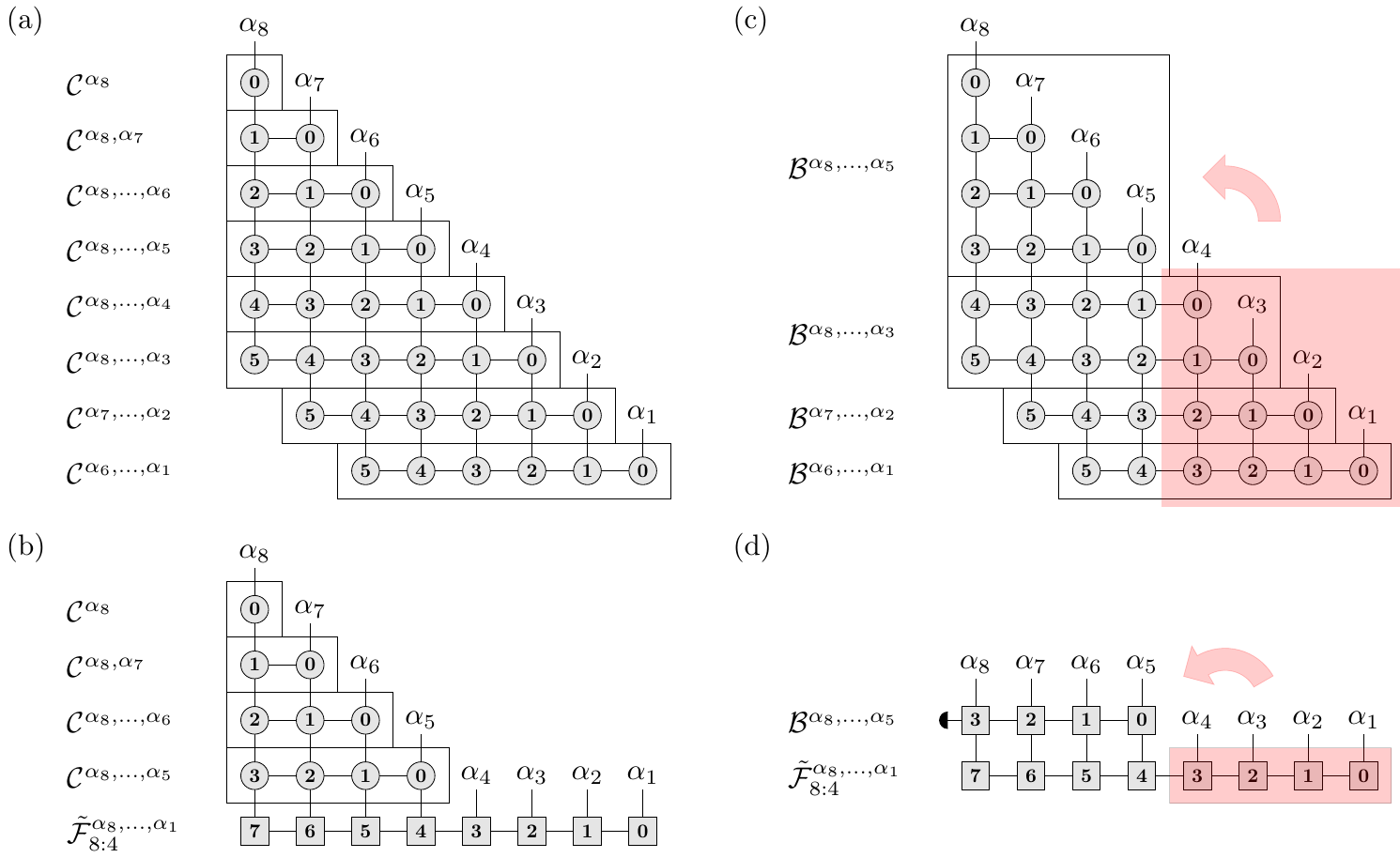}\vspace*{7mm}
\caption{The tensor network for a PT with $n=8$ time steps and memory cutoff 
$n_c=6$ using the sequential algorithm by J{\o}rgensen and Pollock~\cite{JP} 
(a and b) as well as our divide-and-conquer scheme (c and d), respectively.
(a) and (c) show the subdivision of
the overall tensor network into different blocks (black rectangles) 
to be contracted from bottom to top. 
In the sequential algorithm, this is done row by row (b).
The divide-and-conquer algorithm is based on the observation 
that the last block in (c), formed by the topmost $n/2$ lines, 
exactly matches the part of the already contracted network shaded in red.
This enables a contraction of multiple rows at a time as depicted in (d).
The black semicircle seen here represents a ``closure'' that describe tracing out dangling bonds; see text for further discussion.
\label{fig:theory}}
\end{figure*}

Performing the Feynman-Vernon summation in Eq.~\eqref{eq:feynmanvernon} is notoriously 
difficult. For a system Hilbert space of dimension $D$, 
the sum involves $D^{2n}$ terms and, thus, scales
exponentially with the total number of time steps $n$. 
This issue can be addressed~\cite{TEMPO,JP}
by representing the influence functional in a more convenient
form using matrix product operators (MPOs)~\cite{MPS_Orus,MPS_Schollwoeck}.
It is then referred to as the process tensor matrix product operator (PT-MPO)
\begin{align}
\mathcal{F}^{\alpha_n,\dots,\alpha_1}=&
\sum_{d_{n-1}, \dots, d_1} \mathcal{Q}^{\alpha_n}_{1,d_{n-1}}
\mathcal{Q}^{\alpha_{n-1}}_{d_{n-1},d_{n-2}}\dots 
\mathcal{Q}^{\alpha_2}_{d_2,d_1}\mathcal{Q}^{\alpha_1}_{d_1,1},
\label{eq:processtensor}
\end{align}
where the monolithic tensor $\mathcal{F}^{\alpha_n,\dots,\alpha_1}$
is decomposed into a set of smaller elements
$\mathcal{Q}^{\alpha_{l}}_{d_{l},d_{l-1}}$, which are regarded
as matrices with respect to the inner bond indices $d_l$. 
The latter serve the purpose of conveying time-non-local information,
i.e.~memory, over multiple time steps.
The maximal dimension $\chi$ of the inner bonds $d_l$ strongly depends 
on the complexity of the environment~\cite{PT_PRA}.

The PT-MPO representation achieves the reduction of
Eq.~\eqref{eq:feynmanvernon} to
\begin{align}
\label{eq:propagate}
\bar{\rho}_{\alpha_n} =& \sum_{\substack{\alpha_{n-1},\dots,\alpha_0\\
d_{n-1}, \dots, d_1}}
\bigg(\prod_{l=1}^{n}Q^{\alpha_l}_{d_l,d_{l-1}}
\mathcal{M}_{\alpha_l,\alpha_{l-1}} \bigg) \bar{\rho}_{\alpha_0}
\end{align} 
which can be summed sequentially, one time step at a time, 
with the complexity of $n$ matrix multiplications 
of dimension $D^2 \chi$.
The main challenge for simulating the open quantum system is thus to
bring the influence functional in Eq.~\eqref{eq:IF}
into the form of the PT-MPO in Eq.~\eqref{eq:processtensor}.

\subsection{\label{sec:JP}Sequential algorithm}
J{\o}rgensen and Pollock~\cite{JP} devised the algorithm that is currently most commonly 
used to obtain PTs in MPO form for Gaussian environments.
To distinguish it from our divide-and-conquer scheme, we refer to the approach
in Ref.~\cite{JP} as the sequential algorithm.
The common starting point of both algorithms is the graphical representation
of the double product on the right hand side of Eq.~\eqref{eq:IF} 
in the form of a triangular tensor network (introduced in Ref.~\cite{TEMPO}),
depicted in Fig.~\ref{fig:theory}(a) and (c) for $n=8$ time steps and
memory cutoff $n_c=6$. The latter limits the maximum length of 
the rows.

Each node represents a factor 
$c^{\alpha_{l}}_{\beta_{l+1},\beta_l}
=\delta_{\beta_{l+1},\beta_l} \big[b_l\big]^{\alpha_{l+j},\beta_l}$,
where system Liouville space indices $\alpha_l$ are represented by upward 
facing links, while $\beta_{l+1}$ and 
$\beta_l$ are links to the left and right, respectively. Left or right 
dangling indices are traced out.
Note that, in the notation of Ref.~\cite{JP}, the upward facing links are meant to pass by those nodes further up the tensor network, and then connect to the same outer index $\alpha_l$. The more conventional interpretation of tensor networks, where a bottom node is instead connected to its neighbor to the top, is regained by adding a fourth leg with index $\tilde{\alpha}_l$ at the bottom of each node (except those on the bottom row) via another Kronecker delta, i.e. $\tilde{c}^{\alpha_{l},\tilde{\alpha}_l}_{\beta_{l+1},\beta_l}=c^{\alpha_{l}}_{\beta_{l+1},\beta_l} \delta_{\alpha_l,\tilde{\alpha}_l}$.  In the following we however retain the notation of Ref.~\cite{JP}, labeling only three indices on such tensors.

\begin{figure}
\includegraphics[width=0.48\textwidth]{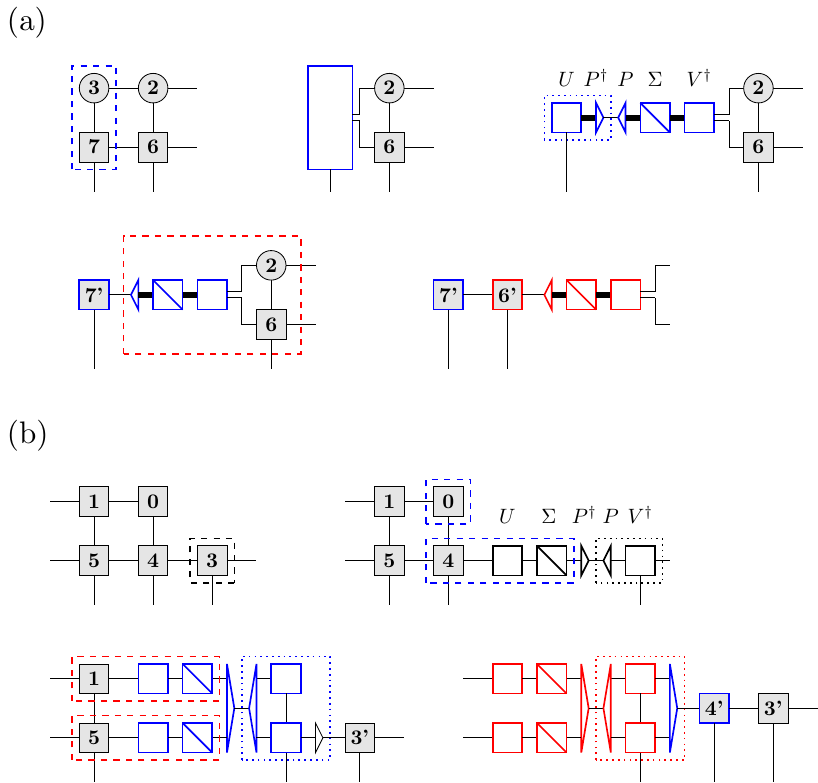}
\caption{\label{fig:compression}
(a): MPO combination and compression by backward sweep used in the sequential algorithm [see Fig.~\ref{fig:theory}(b)]. The units enclosed by the dashed lines are combined into a single matrix. A truncated SVD $U P^\dagger P \Sigma V^\dagger$ is performed, where $P$ denotes the projection onto the space spanned by singular vectors with singular values $\sigma\ge \epsilon \sigma_0$, which is shown as a triangle reducing thick connection (with large dimensions) to thin connection (with smaller dimensions). 
The dotted lines indicate units which are combined to form the nodes in the updated PT-MPO. The part $P\Sigma V^\dagger$ is passed onto the nodes corresponding to the previous time step, before another SVD is performed. This process is repeated until the first time step is reached.
(b): MPO combination and index preselection for the divide-and-conquer method [see Fig.~\ref{fig:theory}(d)]. A forward sweep with truncated SVDs $U \Sigma P^\dagger P V^\dagger$  is performed on the last matrix of the bottom PT-MPO block before it overlaps with matrices of the top PT-MPO block. The corresponding matrices $U \Sigma$ are passed onto the next MPO matrix. SVDs are performed at the top and bottom blocks independently. Singular values from both SVDs are used to identify the projectors onto the subspaces spanned by singular vectors for which $\sigma^{(1)}\sigma^{(2)}\ge \epsilon \sigma^{(1)}_0\sigma^{(2)}_0$, which is shown as a large triangle with three external legs. Updated blocks of the PT-MPO are formed by compressing the Kronecker product of the corresponding matrices $V^{(1)\dagger}$ and $V^{(2)\dagger}$ using these projectors (dotted boxes). These steps are repeated until the last time step is reached.
}
\end{figure}

In the sequential algorithm, the PT is written as a product of rows
\begin{align}
\mathcal{F}^{\alpha_k,\dots,\alpha_1}=&
\prod_{j=1}^k \mathcal{C}^{\alpha_{k},\dots,\alpha_j}
=\prod_{j=1}^k\bigg[ \sum_{\beta_{k-j+1},\dots,\beta_0}
\prod_{l=0}^{k-j}c^{\alpha_{l+j}}_{\beta_{l+1},\beta_l}
\bigg],
\end{align}
where $\mathcal{C}^{\alpha_{k},\dots,\alpha_j}$ is the $j$-th row in the
tensor network, which, as described above, may be reduced to a product 
of at most $n_c$ terms if memory truncation is employed.
The PT $\mathcal{F}^{\alpha_k,\dots,\alpha_1}$ is constructed by multiplying row
after row from bottom to top using the recursion 
visualized in Fig.~\ref{fig:theory}(b),
\begin{align}
\tilde{\mathcal{F}}^{\alpha_k,\dots,\alpha_1}_{(k:j+1)}=&
\mathcal{C}^{\alpha_{k},\dots,\alpha_j}
\tilde{\mathcal{F}}^{\alpha_k,\dots,\alpha_1}_{(k:j)},
\label{eq:JPiter}
\end{align}
with initial value $\tilde{\mathcal{F}}^{\,\alpha_k,\dots,\alpha_1}_{(k:1)}=
\mathcal{C}^{\alpha_{k},\dots,\alpha_1}$.
The final PT is identified as 
$\mathcal{F}^{\alpha_n,\dots,\alpha_1}=
\tilde{\mathcal{F}}^{\alpha_n,\dots,\alpha_1}_{(n:n)}$.
Crucially, the iteration in Eq.~\eqref{eq:JPiter} retains the MPO form
$\tilde{\mathcal{F}}^{\alpha_k, \dots, \alpha_1}_{k:j}
=\sum_{d_k,\dots,d_1} \prod_{l=1}^k \big[\tilde{f}^{\alpha_{l}}_{k:j}\big]_{d_l,d_{l-1}}$ 
with
\begin{align}
\big[\tilde{f}^{\alpha_{l}}_{k:j+1}\big]_{d_l,d_{l-1}}=&
\big[\tilde{f}^{\alpha_{l}}_{k:j}\big]_{d'_l,d'_{l-1}}
\big[c^{\alpha_l}\big]_{\beta_l,\beta_{l-1}}.
\end{align}
However, the dimension of inner indices $d_l = (d'_l, \beta_l)$ is expanded 
to the product of the inner dimension of the previous PT $d'_l$ with that of a
single row $\beta_l$. If $\chi'$ denotes the typical inner dimension of the MPO
$\tilde{\mathcal{F}}^{\,\alpha_k,\dots,\alpha_1}_{(k:j)}$ and the inner dimension
of a single line corresponds to $D^2$ with system Hilbert space dimension
$D$, the inner dimension of the product is now increased to $\chi=\chi' D^2$.
To keep the inner dimensions tractable,
the MPO is compressed by sweeping across it and applying truncated 
singular value decompositions (SVDs) to every element. 
In the forward direction (increasing time step indices), one calculates the
SVD
\begin{align}
\big[\tilde{f}^{\alpha_{l}}_{k:j}\big]_{d_l,d_{l-1}} =&
\sum_{s} U_{d_l, s} \; \sigma_s \; V^\dagger_{s, (\alpha_l,d_{l-1})},
\label{eq:fw_sweep}
\end{align}
with non-negative singular values $\sigma_s$ and matrices $U$ and $V$ with
orthogonal columns.  
Keeping only terms $s$ with significant values $\sigma_s \ge \epsilon \sigma_0$, where 
$\epsilon$ is a given truncation threshold and $\sigma_0$ is the largest
singular value, one replaces
\begin{subequations}\label{eq:fw_sweep_replace}
\begin{align}
\big[\tilde{f}^{\alpha_{l}}_{k:j}\big]_{s,d_{l-1}}  \longleftarrow &
V^\dagger_{s, (\alpha_l,d_{l-1})},
\\
\big[\tilde{f}^{\alpha_{l+1}}_{k:j}\big]_{d_{l+1},s}  \longleftarrow &
\sum_{d_l}\big[\tilde{f}^{\alpha_{l+1}}_{k:j}\big]_{d_{l+1}, d_l} 
U_{d_l, s} \; \sigma_s,
\end{align}
\end{subequations}
which passes on the singular values---indicators for the importance for the 
represented degree of freedom---to the next element in the PT while
simultaneously reducing the inner dimension to the number of significant
singular values. 
A forward sweep is followed by an analogous backward sweep. 
While this scheme to compress the PT-MPO was introduced in Ref.~\cite{JP}, we find empirically that changing the order of sweeps, namely performing the backward sweep before the forward sweep, leads to smaller inner PT dimensions $\chi$ and hence to shorter computation times. For all simulations presented here, we therefore consistently compress PT-MPOs in this reversed order. The combination and backward sweep is diagrammatically represented in Fig.~\ref{fig:compression}(a).

Overall, the calculation of a PT-MPO using the sequential algorithm 
requires $\mathcal{O}(n^2)$ SVDs without memory truncation, or 
$\mathcal{O}(n n_c)$ SVDs with memory truncation.
Exact algorithms to perform SVDs of $N\times M$ matrices with $N>M$ generally require $\mathcal{O}(NM^2)$ floating point operations. Denoting by $\chi'$ the typical inner bond dimension of the PT-MPO after the previous iteration step, the corresponding matrix dimensions for a single SVD as in Eq.~\eqref{eq:fw_sweep} are $M=D^2\chi'$ and $N=D^4\chi'$, respectively, where a factor $D^2$ in the long matrix dimension $N$ stems from the outer bonds of the PT-MPO while the remaining factors originate from combined inner indices. Consequently, the number of floating point operations required per SVD is $\mathcal{O}({\chi'}^3 D^8)$.

\subsection{\label{sec:DnC}Divide-and-conquer algorithm}
Guided by the visual representation of the PT contraction in 
Fig.~\ref{fig:theory}(c), we suggest an alternative algorithm based on the
observation that the triangular tensor network contains a high degree
of self-similarity:
The first and second rows are identical up to a shift. Rows three and four
together form a shifted and truncated replica of the combination of rows one and two. 
Similarly, rows five to eight are the same as the part 
of rows one to four shaded in red in Fig.~\ref{fig:theory}(c). 

Concretely, the tensor network can be contracted in blocks of 
rows, with sizes progressing in powers of two, by iterating:
\begin{align}
\tilde{\mathcal{F}}^{\alpha_k, \dots, \alpha_1}_{k:2^{m+1}}=&
\mathcal{B}^{\alpha_k,\dots, \alpha_{(2^m +1)}}
\tilde{\mathcal{F}}^{\alpha_k, \dots, \alpha_1}_{k:2^{m}},
\label{eq:iter2}
\end{align}
where the block $\mathcal{B}^{\alpha_k,\dots, \alpha_{(2^m +1)}}$ is itself 
contained in the MPO of 
$\tilde{\mathcal{F}}^{\alpha_k, \dots, \alpha_1}_{k:2^{m}}$ and 
can be written as
\begin{align}
\mathcal{B}^{\alpha_k,\dots, \alpha_{(2^m +1)}}=&
\sum_{d_{k-2^m},\dots,d_0} q_{d_{k-2^m}}
\prod_{l=1}^{k-2^m}\big[\tilde{f}^{\alpha_{l}}_{k:2^m}\big]_{d_{l},d_{l-1}}.
\end{align}
Here, $q_{d_{k-2^m}}$, represented as a black semicircle in Fig.~\ref{fig:theory}(d),
define objects we call ``closures'' that describe tracing out dangling bonds.
A similar object has been referred to as a ``cap tensor''               
in other work~\cite{Fux_spinchain}.
Such objects are trivial before  MPO compression, but after compression become non-trivial. 
However, the closures can be calculated iteratively from PT-MPOs by using
trace preservation~\cite{ACE}. For the generalized spin-boson model 
considered here, one starts with $q_{d_{n}=1}=1$ and iterates
$q_{d_{i-1}}=\sum_{d_i} q_{d_{i}}\mathcal{Q}^{\alpha_i=(x,x)}_{d_i, d_{i-1}}$,
where $x$ is a fixed but arbitrary system Hilbert space index (e.g., $x=1$).
As can be seen in Eq.~\eqref{eq:b}, setting $\alpha_i=(x,x)$ 
results in $\big[b_{(i-j)}\big]^{\alpha_i=(x,x),\alpha_j}=1$.
This effectively terminates the double product in Eq.~\eqref{eq:IF} 
at an ealier time step. 

The proposed algorithm follows the divide-and-conquer paradigm because
the total number of rows of the tensor network is divided into an upper and
lower block, where only the lower block has to be calculated explicitly.
The lower block is then again divided into two halves, and so on.
Like other divide-and-conquer schemes, such as the famous 
fast Fourier transformation algorithm~\cite{CooleyTukey, NumericalRecipes},
the numerical complexity is reduced from $\mathcal{O}(n^2)$ to 
$\mathcal{O}(n \log n)$ operations. In our case these operations are SVDs.

Despite the favourable scaling with $n$, the divide-and-conquer algorithm faces some challenges. These arise because of the size of matrices we must combine.
In the sequential algorithm,  each step
combines a MPO with potentially large inner dimension $\chi'$ and a
single line of relatively small inner dimension of $D^2$.
In contrast, in our divide-and-conquer 
algorithm, two MPOs of similar inner dimensions $\chi'$ are combined to $\chi={\chi'}^2$. 
Because of the scaling of SVD routines with the third power of the combined inner dimension together with a factor $D^2$ from the outer bond, the direct application of SVDs would lead to a typical complexity $\mathcal{O}({\chi'}^6 D^2)$ per SVD, which is prohibitively demanding. 
Clearly, a different way to combine two MPOs is needed.

Here, we address this issue by preselecting relevant degrees of freedom:
Say, we wish to update an MPO with matrices 
$\big[\tilde{f}^{\alpha_{l}}_{k:j}\big]_{d'_l,d'_{l-1}}$ by multiplying it
with an MPO with matrices 
$\big[\tilde{g}^{\alpha_l}\big]_{d''_l,d''_{l-1}}$. We first perform 
SVDs on the individual matrices
$\big[\tilde{f}^{\alpha_{l}}_{k:j}\big]_{d'_l,d'_{l-1}}=
\sum_s U^{(1)}_{d'_l, s} \sigma^{(1)}_s V^{(1)\dagger}_{s, (\alpha_l,d'_{l-1})}$
and 
$\big[\tilde{g}^{\alpha_l}\big]_{d''_l,d''_{l-1}}=
\sum_t U^{(2)}_{d''_l, t} \sigma^{(2)}_t V^{(2)\dagger}_{t, (\alpha_l,d''_{l-1})}$,
respectively.
Using this, the matrices of the combined MPO can be formally written as
\begin{align}
&\big[\tilde{f}^{\alpha_{l}}_{k:j+1}\big]_{(d'_l,d''_l),(d'_{l-1},d''_{l-1})}=
\big[\tilde{f}^{\alpha_{l}}_{k:j}\big]_{d'_l,d'_{l-1}}
\big[\tilde{g}^{\alpha_l}\big]_{d''_l,d''_{l-1}}
\nonumber\\&=
\sum_{s,t} 
\big(U^{(1)}_{d'_l, s} U^{(2)}_{d''_l, t} \big)
\sigma^{(1)}_s \sigma^{(2)}_t 
\big(V^{(1)\dagger}_{s, (\alpha_l,d'_{l-1})}V^{(2)\dagger}_{t, (\alpha_l,d''_{l-1})}\big).
\label{eq:select_product}
\end{align}
Note again that, to keep the notation similar to the sequential algorithm in
Ref.~\cite{JP}, Eq.~\eqref{eq:select_product} is considered 
an element-wise product for each value of $\alpha_l$.
In the established notation for more general tensor networks,
Eq.~\eqref{eq:select_product} would be considered a contraction over outer 
indices after including a Kronecker delta
$\big[\tilde{f}^{\alpha_{l}}_{k:j+1}\big]_{(d'_l,d''_l),(d'_{l-1},d''_{l-1})}=
\sum\limits_{\tilde{\alpha}_l}\big[\tilde{f}^{\tilde{\alpha}_{l}}_{k:j}\big]_{d'_l,d'_{l-1}}
\big[\delta_{\tilde{\alpha}_l,\alpha_l}\tilde{g}^{\alpha_l}\big]_{d''_l,d''_{l-1}}$.

To reduce the inner dimension, we keep only combinations of indices $(s,t)$ 
for which the product of singular values
$\sigma^{(1)}_s \sigma^{(2)}_t \ge \epsilon_\textrm{select}
\sigma^{(1)}_0 \sigma^{(2)}_0$
exceeds a value determined by a given threshold $\epsilon_\textrm{select}$.
In practice, the new matrix is directly set to
\begin{align}
\big[\tilde{f}^{\alpha_{l}}_{k:j+1}\big]_{(s,t),(d'_{l-1}, d''_{l-1})}
\longleftarrow 
V^{(1)\dagger}_{s, (\alpha_l,d'_{l-1})}V^{(2)\dagger}_{t, (\alpha_l,d''_{l-1})}
\end{align}
for the subset of $(s,t)$ obeying the condition above,
while the remaining terms are passed on to the next matrices of the 
original MPOs:
\begin{align}
\big[\tilde{f}^{\alpha_{l+1}}_{k:j}\big]_{d'_{l+1},s} \longleftarrow &
\sum_{d'_l} \big[\tilde{f}^{\alpha_{l+1}}_{k:j}\big]_{d'_{l+1},d'_{l}} 
U^{(1)}_{d'_l, s} \sigma^{(1)}_s, \\
\big[\tilde{g}^{\alpha_{l+1}}\big]_{d''_{l+1},t} \longleftarrow &
\sum_{d''_l}\big[\tilde{g}^{\alpha_{l+1}}\big]_{d''_{l+1},d''_{l}} 
U^{(2)}_{d''_l, t}\sigma^{(2)}_t.
\end{align}

The combined forward sweep, selection, and combination process is depicted in Fig.~\ref{fig:compression}(b). It is followed by a backward sweep to further reduce bond dimensions. 
Over a wide range of different examples, we find that the selection reduces the combined inner dimensions from ${\chi'}^2$ to $\lambda\chi'$ with an empirical factor $\lambda$ between $2$ and $12$. The resulting number of floating point operations per SVD $\mathcal{O}(\lambda^3 {\chi'}^3 D^2)$ is therefore nominally comparable to that in the sequential algorithm, $\mathcal{O}({\chi'}^3 D^8)$. However, the massive reduction of inner bonds by our selection strategy comes at the cost of not guaranteeing optimal low-rank approximation. This can result in larger bond dimensions $\chi'$ and therefore in higher numerical demands per SVD compared to the sequential algorithm. 
As discussed in  more detail in Appendix~\ref{app:ratio}, this issue can be ameliorated by choosing different
thresholds for the selection and backward sweep compared to the forward sweep,
which we parameterize by the ratios
${r}_s=\epsilon_\textrm{select}/\epsilon_\textrm{forward}$ and
${r}_b=\epsilon_\textrm{backward}/\epsilon_\textrm{forward}$, 
while at the same time we obtain similar accuracies as for the sequential algorithm 
if we identify $\epsilon_\textrm{forward}$ (the largest of our thresholds) with the nominal threshold $\epsilon$ of the sequential algorithm.

Finally, it is noteworthy that our divide-and-conquer algorithm can be 
implemented in place, i.e. in such a way that only a single copy of the 
full PT-MPO has to be stored, despite the combination of MPO matrices at
different positions in the MPO chain.
This  reduces the  memory footprint of the algorithm.
Furthermore, 
as the line sweeps access the individual MPO matrices in a predictable order,
storing the PT-MPO on hard disk and preloading blocks of matrices can be 
efficiently implemented, with only  a small overhead of about 10\% in 
overall computation time. This enables the calculation of PTs with many
more time steps $n$ than would be possible if required to keep the full MPO in memory.

It should though be noted that, as discussed above, the time required for each SVD scales as $\mathcal{O}(\chi^3)$. Since $\chi$, in general,
depends on the complexity of the environment as well as on the total propagation time in the spirit of the Lieb-Robinson theorem~\cite{Plenio_LiebRobinson}, the actual scaling of the computation time with $n$ may be larger than the scaling $\mathcal{O}(n \log n)$ of the number of SVDs in the divide-and-conquer algorithm.
Benchmarking the numerical run times, as is done in section~\ref{sec:applications}, is therefore essential to assess the overall scaling of the algorithm for typical scenarios in open quantum systems.

\subsection{\label{sec:periodicPT}Finite $n_c$ and Periodic Process Tensors}
\begin{figure}
\includegraphics[width=0.95\linewidth]{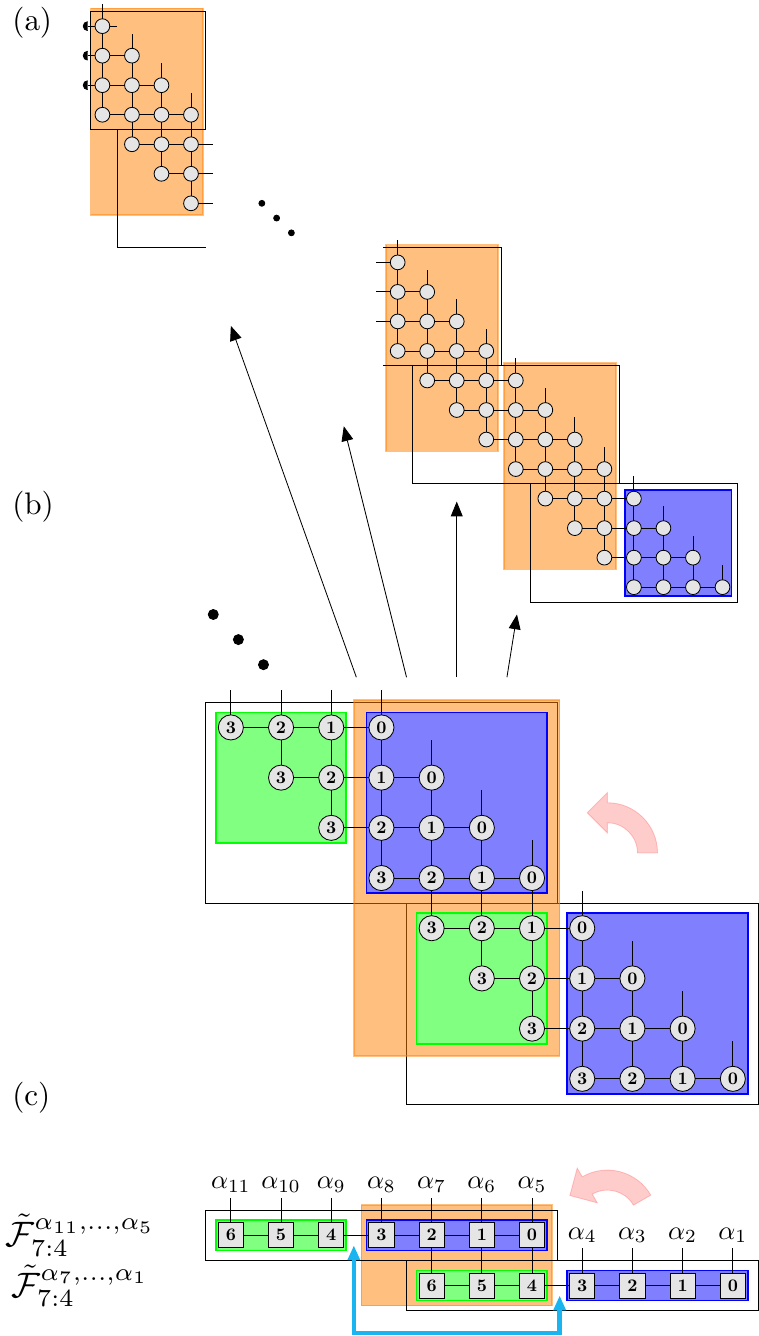}
\caption{(a) A tensor network with many time steps, $n\gg n_c$, is decomposed 
into blocks of length $n_c$. The parts of the tensor network highlighted 
in orange are identical.
(b) The first $2n_c$ rows of the tensor network can be subdivided into 4 blocks, 
where blocks of the same colour are identical. 
The combination of a blue block stacked on top of a green block forms
an orange block, which can be used as a fundamental unit of a periodic PT 
and can then be repeated indefinitely, as depicted in (a). This is due to the fact that 
the left and right interfaces of the orange block, marked by the light blue arrow in (c), seamlessly link together, even after MPO compression of the first $n_c$ rows.
\label{fig:periodic}}
\end{figure}
A useful feature of the sequential algorithm is that it
profits significantly from memory truncation.
If the memory of the environment becomes negligible after $n_c$ time steps,
the number of overlapping columns in subsequent 
rows in the tensor network is limited to $n_c-1$, 
as can be seen, e.g., in the three bottom rows of Fig.~\ref{fig:theory}(a).
Only in the overlap region do MPO compression sweeps have to be performed,
reducing the number of SVDs to $\mathcal{O}(n n_c)$.
This suggests that it is also worthwhile
to investigate how memory truncation can be
incorporated into the divide-and-conquer algorithm.

To this end, consider an intermediate step in the divide-and-conquer 
algorithm where a block of $2n_c$ rows is formed by two blocks of
$n_c$ rows, as depicted in Fig.~\ref{fig:periodic}(b) for $n_c=4$.
In principle, following the same rationale as in the sequential algorithm,
 sweeping only over the finite overlap of $n_c-1$ columns would produce an 
algorithm scaling as $\mathcal{O}\big( n_c \log n\big)$.
Remarkably, for $n_c\ll n$, this means that obtaining the total
PT involves fewer than $n$ SVDs, which implies that many MPO matrices of the 
final PT must be exact copies of others. 

In fact, one can identify a structure in PTs that can be repeated indefinitely.  
Such an observation is analogous to that made in introducing repeating tensors that represent the state of spatially-infinite systems with translational invariance, in algorithms such as
infinite density-matrix renormalisation group (iDMRG)~\cite{Ostlund1995Thermodynamic}, 
infinite time-evolving block decimation (iTEBD)~\cite{Vidal2007Classical}, and infinite projected entangled pair states (iPEPS)~\cite{Jordan2008Classical}.
We now consider how this can be constructed for the periodic PT.
Consider again the extension of the tensor network from $n_c$ to $2n_c$ 
rows. 
For this step, the iteration Eq.~\eqref{eq:iter2} becomes
\begin{align}
\tilde{\mathcal{F}}^{\,\alpha_{3n_c-1}, \dots, \alpha_1}_{3n_c-1:2n_c}=&
\tilde{\mathcal{F}}^{\,\alpha_{3n_c-1}, \dots, \alpha_{n_c+1}}_{2n_c-1:n_c}
\tilde{\mathcal{F}}^{\,\alpha_{2n_c-1}, \dots, \alpha_1}_{2n_c-1:n_c},
\label{eq:iter3}
\end{align}
as shown in Fig.~\ref{fig:periodic}(c).
This has explicit elements
\begin{align}
&\big[\tilde{f}^{\alpha_{l}}_{3n_c-1:2n_c}\big]_{(d_{l},d'_l),(d_{l-1},d'_{l_1})}
 = \nonumber\\
&\begin{cases}
\big[\tilde{f}^{\alpha_{l}}_{2n_c-1:n_c}\big]_{d_{l},d_{l-1}}
\delta_{d'_l,1}\delta_{d'_{l-1},1}, &
l\le n_c,\\
\big[\tilde{f}^{\alpha_{l}}_{2n_c-1:n_c}\big]_{d_{l},d_{l-1}}
\big[\tilde{f}^{\alpha_{l-n_c}}_{2n_c-1:n_c}\big]_{d'_{l},d'_{l-1}}
, & n_c< l \le 2 n_c,  \\
\big[\tilde{f}^{\alpha_{l-n_c}}_{2n_c-1:n_c}\big]_{d'_{l},d'_{l-1}}
\delta_{d_l,1}\delta_{d_{l-1},1}, &
l > 2 n_c,
\end{cases}
\label{eq:periodic_cases}
\end{align}
where we define 
$\big[\tilde{f}^{\alpha_{l}}_{2n_c-1:n_c}\big]_{d_{n_c},d_{n_c-1}}
=\delta_{d_{n_c},d_{n_c-1}}$ to artificially extend the overlap from
$n_c-1$ to $n_c$ columns. 
This is done to achieve a particular partitioning, where 
the first and last cases in Eq.~\eqref{eq:periodic_cases}
correspond to the areas shaded in blue and green, respectively, in
Fig.~\ref{fig:periodic}(b,c). 
If shifted and put together, they exactly reproduce the original block 
$\tilde{\mathcal{F}}^{\,\alpha_{2n_c-1}, \dots, \alpha_1}_{2n_c-1:n_c}$
consisting of the bottom $n_c$ rows.

The role of the middle section in Eq.~\eqref{eq:periodic_cases}, 
visualized as orange box in Fig.~\ref{fig:periodic}(b,c), becomes clear when the
tensor network is further extended by another $n_c$ rows, 
as in Fig.~\ref{fig:periodic}(a). 
The overlap region with the new block of rows again contains
$n_c$ columns, but these are exact copies of the (orange) middle section of 
the MPO, and so do not have to be calculated anew.
Note that the fact that multiple central (orange) blocks fit seamlessly 
together is non-trivial, 
as the inner bonds between the subsequent matrices in the MPO
have been strongly modified by MPO compression, so their relation to the 
bonds in the original tensor network is no longer obvious. 
In particular, inner bonds at different positions within the MPO generally 
have different dimensions. 
Here, however, we have constructed the central (orange) blocks in such a way
that the left and right bonds, as shown by the links marked by the light blue arrow in Fig.~\ref{fig:periodic}(c), correspond exactly to the bonds between
the first (blue) and last (green) section in Eq.~\eqref{eq:periodic_cases}. 
This allows us to repeat the orange block indefinitely and we arrive 
at the periodic PT
$\mathcal{F}^{\,\alpha_{n}, \dots, \alpha_1}= \sum_{d_n, \dots d_1}  
q_{d_n} \prod_{l=1}^{n}\mathcal{Q}^{\alpha_l}_{d_l,d_{l-1}}$
with 
\begin{align}
\label{eq:Qrepeat}
&\mathcal{Q}^{\alpha_l}_{(d'_{l},d''_l),(d'_{l-1},d''_{l_1})}
 = \nonumber\\
&\begin{cases}
\big[\tilde{f}^{\alpha_{l}}_{2n_c-1:n_c}\big]_{d'_{l},d'_{l-1}}
\delta_{d''_l,1}\delta_{d''_{l-1},1}, &
l\le n_c,\\
\big[\tilde{f}^{\alpha_{(l\textrm{ mod }2n_c)}}_{2n_c-1:n_c}\big]_{d'_{l},d'_{l-1}}
\big[\tilde{f}^{\alpha_{((l-n_c)\textrm{ mod }2n_c)}}_{2n_c-1:n_c}\big]_{d''_{l},d''_{l-1}}
, & l> n_c,\\
\end{cases}
\end{align}
where again the matrices
$\big[\tilde{f}^{\alpha_{l}}_{2n_c-1:n_c}\big]_{d'_{l},d'_{l-1}}$
are those of the MPO
$\tilde{\mathcal{F}}^{\,\alpha_{2n_c-1}, \dots, \alpha_1}_{2n_c-1:n_c}$
describing the bottom $n_c$ lines of the tensor network.
In practice, it is useful to compress the periodic part of the PT-MPO 
once more in a final step using our preselection scheme for combining MPOs with 
large inner dimensions, described in Eq.~\eqref{eq:select_product}.  In doing this, care must
be taken not to modify the left and right interfaces. 

To summarize this section, if the memory of a Gaussian environment is finite, a periodic PT 
can be obtained using only $\mathcal{O}(n_c \log n_c)$ 
rank-reducing SVDs. Remarkably, this is constant in the overall propagation
time $t_e=n\Delta t$ and provides again a nominal scaling advantage over the
sequential algorithm with $\mathcal{O}(nn_c)$. 
A further advantage of the periodic PT is that the memory requirements
for storing it is $\mathcal{O}(n_c)$, 
much smaller than that for storing a full PT $\mathcal{O}(n)$.

In the following sections we will showcase the power of our divide-and-conquer scheme
with and without memory truncation on a series of applications.

\section{\label{sec:applications}Applications}
\subsection{\label{sec:QDPhonon}Fluorescence Spectra of Quantum Dots}
\begin{figure*}
\centering
\includegraphics[width=\textwidth]{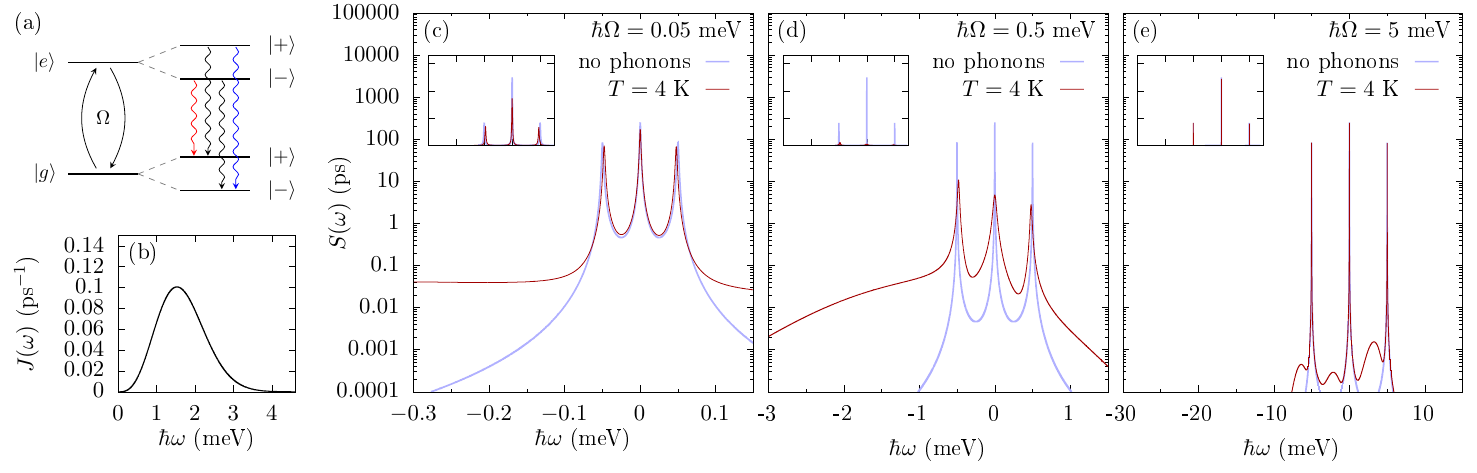}
\caption{\label{fig:QDPhonon}
(a) Cartoon of fluorescence experiment: The external laser drive dresses ground and excited state of the QD. In the absence of a phonon environment, photon emission can be understood in terms of optical transitions between laser-dressed states. Black arrows indicate transitions contributing to photons at the central frequency of the two-level system, whereas red and blue arrows depict transitions detuned by $-\Omega$ and $\Omega$ from the central frequency, respectively.
(b) Superohmic phonon spectral density of a QD with electron radius $a_e=4$~nm.
(c)--(e) Fluorescence spectra of driven QDs with different driving strengths $\Omega$, shown on a logarithmic intensity scale. Insets depict the same spectra on a linear scale. The simulations were performed over $n=2^{21}$ time steps of size $\Delta t=0.01$ ps using periodic PTs with memory cut-off at $n_c=2^{11}$ time steps.
}
\end{figure*}

\subsubsection{Model and context}
Self-assembled III-V semiconductor quantum dots (QDs) 
are key elements for photonic quantum technologies~\cite{Lodahl_QDnetworks}.
Due to their strong interaction with light, they can be used as bright
sources of pure single photons~\cite{PI_singlephoton,Thomas_singlephoton},
entangled photon pairs~\cite{PI_entangled_PRL, Stevenson_entangled}, and other 
non-classical multi-photon states~\cite{Munoz_photon_bundle,PI_Fock}. 
However, electronic excitations in QDs also strongly interact with 
longitudinal acoustic phonons. 
The resulting non-Markovian effects are significant, and so typically cannot be fully
described by weak-coupling master equations~\cite{variationalPRB}.
While polaron master equations accurately account for phonon effects 
in the limit of weak driving~\cite{Review_Nazir}, 
these break down under strong-driving conditions~\cite{variationalPRB}. 
Therefore, numerically exact path integral 
techniques based on QUAPI~\cite{QUAPI1,QUAPI2}
have been applied~\cite{PI_cavityfeeding,PI_singlephoton} 
to investigate the dynamics of driven QDs.

While path integral techniques have allowed calculations of some features of quantum dots,  numerically exact calculation of the spectra of strongly driven QDs is difficult.  Multi-time 
correlations (as needed to obtain emission spectra) can be calculated with such approaches~\cite{PI_multitime}, but reaching convergence remains a challenge.
This is due to a separation of timescales (see Appendix~\ref{app:spectra_convergence} for a convergence study):
Because of the long radiative lifetimes in QDs, the
dynamics has to be propagated to a few nanoseconds in order to avoid artifacts in the Fourier-transformed spectrum.
Typical phonon memory times span several picoseconds.
When considering driving with
high laser intensities, or strongly off-resonant driving~\cite{SUPER},
the resulting  oscillatory dynamics
makes it necessary to choose small time steps $\Delta t$ of the order of
a few femtoseconds.
Taken together, these separate timescales yield a problem where both the number of timesteps $n$, and the memory cutoff $n_c$ must be large to produce accurate results.
Moreover, to analyze phonon sidebands, spectra are often presented on 
a logarithmic scale, which makes numerical errors quite prominent.
One should therefore use very small MPO compression thresholds in 
PT simulations, resulting in sizeable inner dimensions $\chi$.
Nevertheless, for the reasons explained above, this challenging problem can be addressed using our divide-and-conquer scheme, in particular when combined with the use of a periodic PT.

In Fig.~\ref{fig:QDPhonon}, we present the fluorescence spectra of a QD
driven with Rabi frequency $\Omega$ and coupled to a bath of phonons, showing how the Mollow triplet~\cite{Mollow} is affected by this bath.
The phonon-free system evolution with driving and radiative decay with
rate $\kappa$ is given by the Lindblad master equation
\begin{align}
\frac{\partial}{\partial t}\rho=\mathcal{L}_S {\rho} =
\frac i\hbar \big[H_S, \rho\big]
+\kappa\mathcal{D}_{\sigma^-} (\rho) 
\label{eq:MEatom}
\end{align}
with system Hamiltonian
$H_S = \frac{\hbar}{2} \Omega \sigma^x$
and Lindbladian
\begin{align}
\mathcal{D}_{{A}} (\rho) =
{A}\rho{A}^\dagger-\frac 12\big(A^\dagger A\rho + \rho A^\dagger A \big).
\end{align}
Here, the QD is modelled as a two-level system with ground and exciton states
$|g\rangle$ and $|e\rangle$, respectively, and we use the conventional 
notation for operators $\sigma^+=|e\rangle\langle g|$, 
$\sigma^-=|g\rangle\langle e|$, and $\sigma^x=\sigma^+ + \sigma^-$.
The spectrum is obtained as the Fourier transform of the 
two-time correlations:
\begin{align}
S(\omega)=&\lim_{t\to\infty} \lim_{\tau_\infty\to\infty}\textrm{Re}\bigg\{
\int\limits_{0}^{\infty}d\tau\,\Big[
\langle \sigma^+(t+\tau) \sigma^- (t)\rangle 
\nonumber\\&
-\langle \sigma^+(t+\tau_\infty) \sigma^- (t)\rangle 
\Big]e^{-i\omega \tau}\bigg\},
\end{align}
where the coherent scattering contribution (elastic peak) has been subtracted.

For comparison, the spectra for simulations without phonons---as would be relevant for a driven atom---are shown as light blue lines in 
Fig.~\ref{fig:QDPhonon}(c--e) for different Rabi frequencies $\Omega$ and
fixed radiative decay rate $\kappa=(0.5\textrm{ ns})^{-1}$. 
In this case one sees the characteristic Mollow triplet~\cite{Mollow}
with a central peak at the two-level transition frequency and two
side peaks at frequencies $\pm\Omega$. 
The ratio between the heights of the central peak and each side peak is 3:1, while the ratio between the integrated areas is 2:1~\cite{KimbleMandel}.
The shape of the atomic Mollow triplet is explained in terms of
laser-dressed states, i.e., eigenstates 
$|\pm\rangle=\big(|g\rangle \pm |e\rangle\big)/\sqrt{2}$ 
of the Hamiltonian $H_S$ with eigenvalues $\pm\hbar\Omega/2$.
Optical transitions $|-\rangle \to |-\rangle$ and $|+\rangle\to|+\rangle$ 
both contribute to the central peak at the two-level transition frequency,
while transitions $|+\rangle \to |-\rangle$ and $|-\rangle \to |+\rangle$ 
are responsible for the side peaks at $\Omega$ and $-\Omega$, respectively.

\subsubsection{Simulation results and discussion}
In solid-state QDs, emission spectra are strongly affected by interactions 
with phonons.   
The QD-phonon interaction is of the form of a spin-boson model, corresponding to
Eq.~\eqref{eq:spinboson} with a coupling operator 
$\hat{O}=|e\rangle\langle e|$.
The phonon spectral density takes the superohmic form:
\begin{align}
J(\omega)=\frac{\omega^3}{4\pi^2\rho\hbar c_s^5}
\bigg(D_e e^{-\omega^2a_e^2/(4c_s^2)} - D_he^{-\omega^2a_h^2/(4c_s^2)}\bigg)^2.
\end{align}
For a GaAs-based QD, we take parameters given in  Ref.~\cite{Krummheuer} corresponding to an electron and hole radii $a_e=4$ nm and $a_h=a_e/1.15$, respectively, and we show the resulting spectral density in
Fig.~\ref{fig:QDPhonon}(b). The corresponding polaron shift or reorganization
energy $\int\limits_0^\infty d\omega J(\omega)/\omega\approx 0.072$ meV 
is absorbed into the definition of the excited state energy.
The remaining physical and convergence parameters  for our simulations 
using divide-and-conquer combined with periodic PTs are
the initial phonon temperature $T=4$~K, time discretization 
$\Delta t=0.01$ ps, memory time $t_\textrm{mem}=n_c \Delta t$ with 
$n_c=2\,048$, 
total propagation time $t_e=n \Delta t$ with $n=2^{21}\approx 2\times 10^6$, and
MPO compression threshold $\epsilon=10^{-12}$.

Previous works have addressed QD fluorescence spectra in the regime of 
weak Rabi driving using polaron master 
equations~\cite{Hughes_polaron2012, Mollow_Nazir}, or path-integral calculations~\cite{PI_multitime}. There, two effects have
been identified:
First, compared to the phonon-free spectra, simulations with phonons reveal
a line broadening, which also reduces the heights of the peaks. 
Second, the side peaks are shifted towards the center 
due to phonon renormalization of the transition dipole.
These effects can be seen in Fig.~\ref{fig:QDPhonon}(c), where we show the weak-driving case of $\hbar\Omega=0.05$ meV.

Importantly, the ability of our divide-and-conquer algorithm to treat problems with many timesteps  enables us to also investigate the regime of stronger
driving, beyond the range of previous work. In Fig.~\ref{fig:QDPhonon}(d), we see that on increasing the driving strength to $\hbar\Omega=0.5$~meV, three changes occur:
the spectral lines are broadened,  an asymmetric background arises, and there is a notable change in the relative height of the three Mollow peaks with the low-energy side peak now dominating the emission.   We discuss each of these features in turn.
The broadening seen is consistent with behavior known in the weak-coupling limit~\cite{Review_Nazir}, 
where the linewidth is proportional to the spectral density $J(\omega)$ 
evaluated at the Rabi frequency $\omega=\Omega$.  Since $J(\Omega)$  is 
significant at $\hbar\Omega=0.5$~meV---see Fig.~\ref{fig:QDPhonon}(b)---the broadening here becomes large.
Regarding the asymmetric background, this feature is similar to that seen in the phonon side band in an undriven QD~\cite{PI_QRT}.
Finally, the  relative peak heights can be understood as the result of fast thermalization in the dressed-state basis.
Thermalization predominantly occupies the lower dressed state $|-\rangle$, 
because the energetic splitting between $|+\rangle$ and $|-\rangle$ 
is larger than the thermal energy $\hbar\Omega> k_B T\approx 0.34$ meV. 
Hence, emission from $|+\rangle$ is quenched. 
Moreover, the direct transition $|-\rangle \to |-\rangle$ emitting
photons at the frequency of the central peak also competes with 
phonon-assisted photoemission processes, where an energy $\hbar\Omega$ is
efficiently absorbed into the phonon bath. The photon with the
remaining energy contributes instead to the lower energy side peak.

Considering even larger driving with $\hbar\Omega=5$ meV, as 
depicted in Fig.~\ref{fig:QDPhonon}(e),  
we find a restoration of the characteristic Mollow triplet with 3:1 ratio between the heights of central
and side peaks just like in the phonon-free simulations.
This is due to dynamical decoupling from 
phonons~\cite{Vagov_nonmonotonic, Kaldewey2017, Denning_decoupling}, which occurs because the
spectral density $J(\omega)$ becomes small for frequencies 
$\omega\gtrsim 5$ meV$/\hbar$, see Fig~\ref{fig:QDPhonon}(b).

\begin{figure}
\centering
\includegraphics[width=\linewidth]{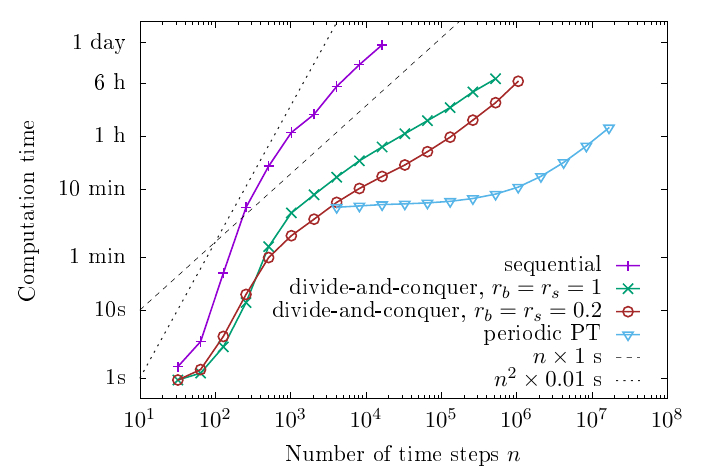}
\caption{\label{fig:QDPhonon_times}
Computation time as a function of the total number of simulation time steps $n$,
on a double-logarithmic scale. Results are shown for the sequential algorithm,
the divide-and-conquer scheme 
with different ratios $r_b=r_s$ between 
compression thresholds for backward and forward sweeps,
and the periodic PT approach starting from a divide-and-conquer calculation with
$r_b=r_s=0.2$. For reference, linear and quadratic scaling with $n$ are 
depicted as dashed dark gray and dotted light gray lines, respectively.}
\end{figure}
\subsubsection{Scaling of computation time}
We next discuss the computational cost of these calculations for the 
various different algorithms presented above.  This is shown in 
Fig.~\ref{fig:QDPhonon_times} where we show the computation time, defined as the total elapsed time 
from start to end of the program on a conventional laptop computer 
with Intel Core i5-8265U processor.

For the sequential algorithm by J{\o}rgensen and Pollock~\cite{JP}, 
we initially see the expected superlinear increase in computation time up to about 
$n\sim 1\,000$ time steps, with a scaling compatible with the 
$\mathcal{O}(n^2)$ behavior of the number of SVDs in the algorithm
(cf.~dotted and dashed lines indicating slopes corresponding to 
$\mathcal{O}(n^2)$ and $\mathcal{O}(n)$ scaling, respectively).
For larger $n$, where $n>n_c=2\,048$, the scaling switches to linear, which again matches the expectation of $\mathcal{O}(n n_c)$ SVDs . The sequential algorithm data is limited to $n\lesssim 10\,000$ timesteps due to the computation time required beyond this.

The divide-and-conquer scheme similarly shows an initial superlinear scaling for 
$n\ll n_c$, but is approximately one order of magnitude faster than the
sequential algorithm after $n\sim 100$ time steps. 
After the kink at $n\sim n_c$, this algorithm scales more slowly, 
with a slope on double-logarithmic scale that indicates sublinear behavior. 

As discussed in the theory section, it can be beneficial
to use different compression thresholds for
selection, backward, and forward sweeps. The computation time for simulations with threshold ratio $r_b=r_s=0.2$ is also shown in the Figure, and is seen to be a factor of three smaller than that with $r_b=r_s=1$. 
We have checked that  this change corresponds to a reduction of the maximal inner bond dimension of the final PT from $\chi=158$ to $\chi=133$.
When studying accuracy vs the forward compression threshold $\epsilon$, we find similar accuracies for both values of threshold ratios.
These calculations allow us to reach $n\sim 1\,000\, 000$ timesteps;  the ultimate limitation in this case is the storage of the full PT-MPO, which requires over 400 Gigabytes, rather than the computation time.

Considering the periodic PT algorithm, we show results from
switching to periodic PTs after calculating an intermediate PT of
$n_c$ rows using the divide-and-conquer scheme with $r_b=r_s=0.2$.
As anticipated, in this case the computation time remains constant until about $n\sim 1\,000\, 000$ 
time steps, because no more numerical resources have to be spent on 
PT calculation. 
At extremely large $n\sim 10^7$, the time becomes dominated by the propagation
of the open quantum system in Eq.~\eqref{eq:propagate}, where PT-MPO is
contracted with the set of system propagators $\mathcal{M}_{\alpha_l, \alpha_{l-1}}$ via the outer indices, and by the extraction of the observables.  
Even though this only involves matrix multiplications, which
are much less demanding than SVDs, they have to be applied at each time
step, and so linear scaling arises for extremely large $n$.
The prefactor of this linear scaling is however far smaller than that of the other algorithms, as is clear from the vertical offset on the double-logarithmic graph. 
Storing the periodic PT requires less than 2 Gigabytes, which elimitates the need for writing to and reading from hard disk.

Finally, we want to point out that there are occasions where PT-MPOs can be employed with genuine sublinear scaling with respect to $n$. For example, if one is interested in the stationary states of an open quantum system with time-independent or periodic system Hamiltonian, one can contract the periodic PT with the corresponding system propagators. Thereby, one obtains an effective propagator for the system expanded by the inner dimension of the PT-MPO at the periodic cut, which can be diagonalized. The stationary state(s) are then given by the eigenvector(s) of the effective propagator corresponding to eigenvalue(s) with value 1.  In addition, multi-time correlations can be calculated directly from considering the other eigenvalues and eigenvectors.

\subsection{\label{sec:superrad}Superradiance}
\begin{figure*}
\centering
\includegraphics[width=\textwidth]{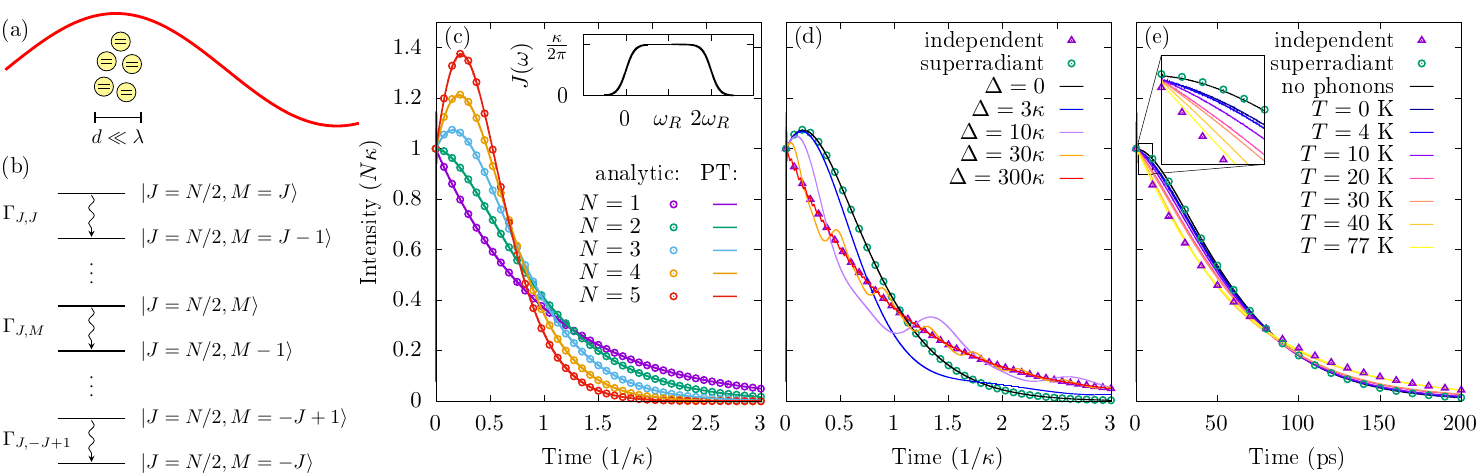}
\caption{\label{fig:superrad}
(a)~Cartoon showing $N=5$ two-level quantum emitters confined to a region much smaller than the wavelength of the emitted light. If the emitters are also spectrally indistinguishable, they effectively couple to the electromagnetic field as a single emitter, resulting in superradiant emission. 
(b)~Superradiant transitions between states in the Dicke ladder.
(c)~Emitted intensity vs time for $N=1\ldots5$ superradiant identical emitters. Points show 
analytic results of rate equations (see text), lines represent divide-and-conquer PT simulations without memory truncation. 
The inset shows the spectral density used to model radiative decay 
at rate $\kappa$. This is flat around the  
central frequency $\omega_R$ with smooth edges.
(d) Breakdown of superradiance for $N=3$ emitters upon increasing spectral
distinguishability. The transition frequencies of the emitters are detuned 
by $-\Delta/2$, $0$, and $\Delta/2$, respectively, from the central frequency.
(e) Effect of phonon baths on superradiance for $N=2$ degenerate quantum dots, for a variety of temperatures  $T$.
}
\end{figure*}

\subsubsection{Model and context}

Superradiance is a dramatic consequence of collective quantum behavior~\cite{Dicke, Haroche}, where $N$ emitters act as one, resulting in spontaneous emission rates 
that scale superextensively with $N$. 
It occurs when the inter-emitter distances are much smaller than the wavelength of the emitted light, making the emitters spatially indistinguishable [see Fig.~\ref{fig:superrad}(a)], and when the energies of emitters are the same, making them spectrally indistinguishable.
In this case, if all emitters are prepared in their excited
state at $t=0$, the emitted intensity features a characteristic burst at 
short times and a slowly decaying tail on long times, in distinction to the exponential
decay that occurs for independent or distinguishable emitters~\cite{Haroche}.
Concomitant with superradiance is the existence of subradiant states, which are optically
inactive and so can be used to store excitations or quantum information
over long periods of time.
Due to recent advances in fabrication, it has become possible to 
explore collective effects of a few spectrally indistinguishable 
solid-state QDs, including coupling multiple QDs into the same photonic
waveguide~\cite{CoopSciAdv,Gammon_superradiance,Waks_superradiance,Lodahl_superradiance}. 
As in the previous section, consideration of semiconductor QDs raises the issue of phonon effects, and a key theoretical question is thus modeling of 
imperfections and environmental effects in real-world
solid-state devices~\cite{CoopWiercinski}.

The standard model to consider for superradiance is $N$ two-level systems coupled to the electromagnetic field via electric dipole interaction.  We can thus write the full Hamiltonian (i.e. $H_S+H_E$) as~\cite{Dicke,Haroche}
\begin{align}
\label{eq:Hsup_orig}
H=& \sum_{j=1}^N \hbar \omega_j \sigma^+_j \sigma^-_j 
+\sum_k \hbar \omega_k a^\dagger_k a_k 
\nonumber\\&
+\sum_{j,k} \hbar \big( g_{j,k} a^\dagger_k +g_{j,k}^* a_k\big) 
\big(\sigma^-_j + \sigma^+_j \big),
\end{align}
where $\omega_j$ is the transition frequency of emitter $j$, $\omega_k$
is the frequency of photon mode $k$, $g_{j,k}$ are the light-matter coupling
constants, $a^\dagger_k$ and $a_k$ create and destroy a photon
in mode $k$, and $\sigma^+_j$ and $\sigma^-_j$ describe the excitation and
deexcitation of the $j$-th emitter.
Throughout this example, we assume $g_{j,k}=g_k$ is real and independent of $j$, which implies that all emitters couple to the photon environment with the same phase, as is crucial for superradiance~\cite{Dicke,Haroche}.

The Hamiltonian written above is a multimode generalization of the Dicke model, without any rotating wave approximation.
As such, the time evolution of this model will involve high frequencies, associated with the characteristic frequencies $\omega_j, \omega_k$.
In many cases, it is preferable to make a rotating wave approximation (RWA), yielding the Tavis--Cummings model, which would then allow one to shift to a frame in which these high frequencies are eliminated.
However, doing so changes the system-environment coupling in a way that precludes the direct application of the PT-MPO formalism discussed above:  After the RWA, the system-environment coupling no longer takes the simple product form in Eq.~\eqref{eq:spinboson}, and instead involves two different system operators coupling to the environment.  
While approaches do exist to construct tensor network representations for the evolution of such problems~\cite{StrathearnBook}, they involve tensors with larger internal dimensions.
However, our divide-and-conquer scheme enables us to resolve very many time steps. 
As such, we are able to calculate a PT-MPO for the photonic environment with the original electric dipole coupling in Eq.~\eqref{eq:Hsup_orig} without making the RWA. 
In appendix~\ref{app:revrwa}, we discuss an alternative perspective, starting from the Tavis--Cummings model (with the RWA), and re-introducing counter-rotating terms as an approximation, controlled by adding a common carrier frequency, $\omega_R$, to the system frequencies $\omega_j,\omega_k$.

\subsubsection{Comparison of numerical and analytic results}

Starting from  Eq.~\eqref{eq:Hsup_orig}, we directly apply our algorithm to the environment Hamiltonian $H_E=H-H_S$ with system Hamiltonian $H_S=\sum_j \omega_j \sigma^+_j\sigma^-_j$.
The form of the light-matter coupling is advantageous,  as degeneracies allow us to treat relatively large $N$ exactly.
Specifically $\sum_j \sigma^x_j$ has a spectrum with 
only $N+1$ different eigenvalues $-N,-N-2,\dots, N$.
This allows us to construct PTs with only $(N+1)^2$ instead of $2^{2N}$ 
different outer indices $\alpha$~\cite{PI_cavityfeeding,TEMPO}.
As a result, the PT-MPO calculation is so efficient that it is the final contraction---describing the propagation of the $N$-emitter system in a $2^{2N}$ dimensional space---rather than the PT calculation that
currently limits the number of emitters $N$ in our numerically exact simulations.

The spectral density used for the PT-MPO calculation, shown in the inset to Fig.~\ref{fig:superrad}(c), is chosen as:
\begin{align}
J(\omega)=\frac{\kappa}{2\pi} \frac{1}{1+e^{-\omega/w}} 
\frac{1}{1+e^{(2\omega_R-\omega)/w}}.
\end{align}
This expression has a wide flat region around $\omega\approx\omega_R$ with a constant
value $\kappa/(2\pi)$, so that in the Markov limit a single emitter 
would spontaneously decay with rate $\kappa$. The edges of this box-shaped spectral density are rounded using logistic functions centered at
$\omega=0$ and $\omega=2\omega_R$, respectively, with transition
widths $w=\omega_R/10$. We find such rounding produces PT-MPOs with smaller inner dimensions than for a sharp cutoff.
Throughout this discussion, we choose 
$\omega_R=1000\kappa$, time steps $\Delta t=1/(8192\kappa)$, total number
of time steps $n=t_e/\Delta t=32\,768$, and MPO compression threshold 
$\epsilon=10^{-9}$. The divide-and-conquer algorithm is applied without 
memory truncation.
The emitted intensity $I(t)=-\frac{\partial}{\partial t} \sum_j n_j(t)$ is extracted using central differences \mbox{$I(t)\approx -\sum_j [n_j(t+\Delta t)-n_j(t-\Delta t)]/(2\Delta t)$}, where $n_j(t)$ denotes the occupation of the $j$-th emitter.  The results of such calculations for identical emitters, with $\omega_j=\omega_R$, are shown by solid lines in Fig.~\ref{fig:superrad}(c).

For comparison to these simulations, Fig.~\ref{fig:superrad} also shows analytic results.  
These are found using the standard approach of considering transitions between the ladder of Dicke states, as depicted in Fig.~\ref{fig:superrad}(b), and considering the result in the RWA (i.e. Tavis--Cummings model).
The validity of the RWA is controlled by the size of $\omega_R$ compared to other parameters.
The Dicke states $|J,M\rangle$ are  eigenstates  of the collective angular momentum operator $\hat{J}_{x/y/z}=\sum_{j=1}^{N}\sigma^{x/y/z}_j$, specifically
the total angular momentum $\hat{J}^2|J,M\rangle=J(J+1)|J,M\rangle$ and its $z$-component $\hat{J}_z|J,M\rangle=M|J,M\rangle$.
Starting from the fully excited state, $|J=N/2,M=N/2\rangle$, the dynamics is restricted to the states with $J=N/2$ and $M=-J,\dots,J$.
The light-matter interaction involves the collective lowering operator $\hat{J}^-$ which gives rise to optical transitions between states $|J,M\rangle$ to $|J,M-1\rangle$. 
The rates $\Gamma_{J,M}$ for these transitions can be found using Fermi's golden rule, which yields $\Gamma_{J,M}=\kappa(J+M)(J-M+1)$~\cite{Haroche}.
The sequential photon emission processes can be solved analytically; the
solutions are listed explicitly in Appendix~\ref{app:superrad_explicit} and are shown by points in Fig.~\ref{fig:superrad}(c).

We see in Fig.~\ref{fig:superrad}(c) that the emitted intensities from $N=1,\dots,5$ indistinguishable ($\omega_j=\omega_R$ for all $j$) quantum emitters obtained from the PT-MPO simulations very closely match the analytical results.
As noted above, the PT-MPO calculations are for the Dicke model (without RWA), while the analytic results are for the Tavis--Cummings model, with the RWA in the light-matter coupling;  this confirms the RWA would be valid for these parameters.
While not clearly visible in the figure, there are discrepancies between the two calculations at early times, $t\lesssim 1/\omega_R$.  These differences are due to the difference of models with and without the RWA:  there are transient effects of turning on the matter-light coupling including the counter-rotating terms at $t=0$. 
We have checked that, as expected, such effects can be reduced by increasing $\omega_R$ (not shown).

\subsubsection{Effects of distinguishability and decoherence}

While the results above demonstrate the ability of the PT-MPO approach to recover superradiant dynamics for the Dicke model, the fact that all results were recoverable from analytic calculations suggests such an approach was not needed.
However the PT-MPO approach includes the full description of environment influences and so can be used in situations where no analytic solutions are available.  Here we use this to present results on how superradiance is destroyed by spectral distinguishability and  decoherence.

Figure~\ref{fig:superrad}(d) shows the breakdown of superradiance for $N=3$
quantum emitters when they become spectrally distinguishable. 
This is implemented by changing the system Hamiltonian by
setting different values for the two-level transition frequencies 
$\omega_1=\omega_R-\Delta/2$, $\omega_2=\omega_R$, $\omega_3=\omega_R+\Delta/2$.  
We note that this can be done with a single calculation of the PT-MPO, and just changing the system propagator with which it is then contracted.
Indeed, with increasing frequency detuning between emitters, $\Delta$,
the superradiant intensity burst is suppressed, and replaced by oscillations around an exponentially decaying intensity.   At sufficiently large $\Delta$ one recovers mono\-exponential decay with rate $\kappa$ as is expected for  independent emitters.

The PT-MPO approach also allows us to investigate how superradiance
is affected when the emitters are additionally coupled to other environments,
such as phonons. 
These can affect the emission dynamics by dephasing inter-emitter
coherences and by accumulating which-path information, i.e., which of the 
emitters is excited and which is not. In Fig.~\ref{fig:superrad}(e), we show
the corresponding intensity for $N=2$ spectrally indistinguishable 
semiconductor QDs ($\omega_j=\omega_R$) , with each QD coupled to a bath of longitudinal acoustic phonons.
For this bath we use the same spectral density and parameters as in Fig.~\ref{fig:QDPhonon};  we thus present results in physics units, and so require a specific choice of $\kappa=1/64$ ps$^{-1}$.
We find that superradiant emission is already slightly suppressed due to
QD-phonon interactions for baths at initial temperature $T=0$~K; similar behavior persists up to temperatures of $T=4$~K. 
Signatures of superradiance start to significantly 
decrease from about $T=10$~K and almost vanish at liquid Nitrogen 
temperature $T=77$~K.

\subsection{\label{sec:coherence_decay}Coherence decay with a strongly-peaked spectral density}
\begin{figure*}
\centering
\includegraphics[width=\textwidth]{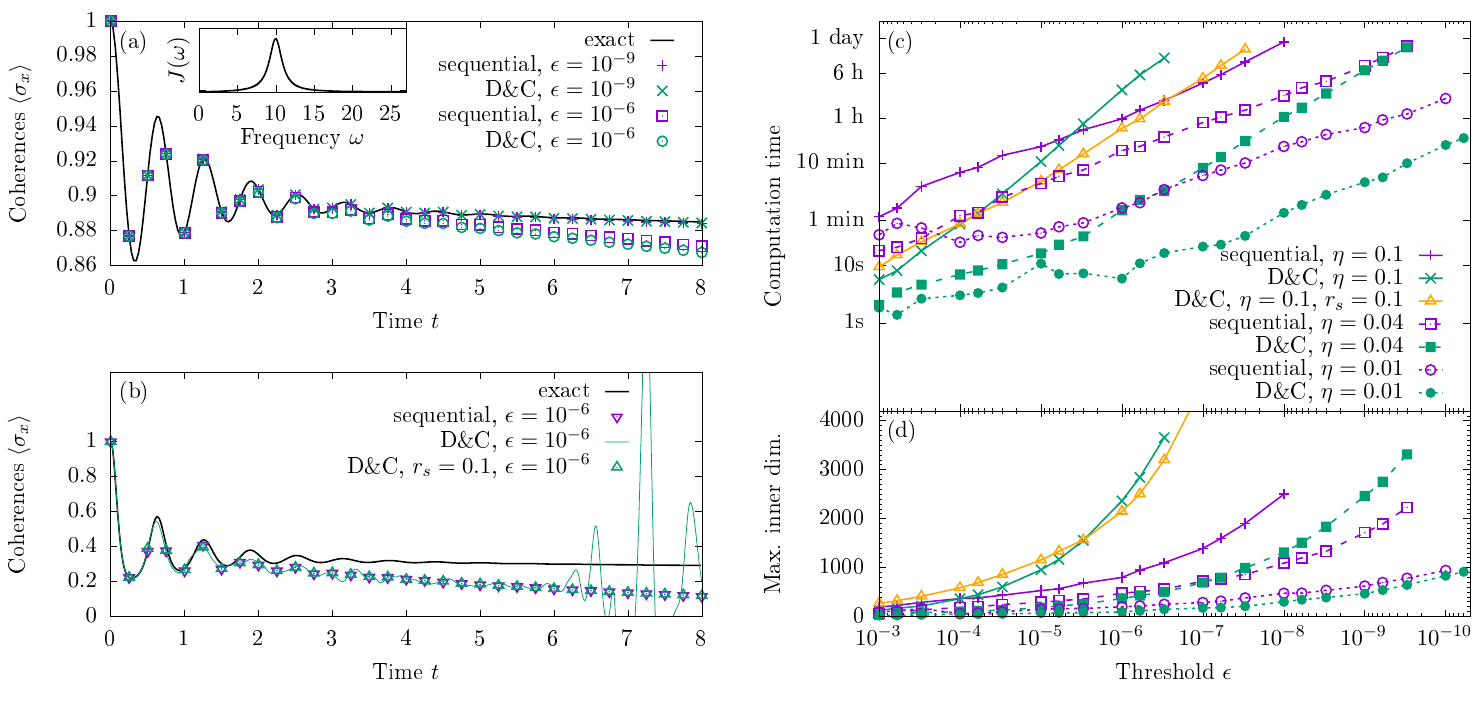}
\caption{\label{fig:coherence_decay}
(a) Free decay of coherences in a two-level system with spectral density $J_{QDT}(\omega)$ [cf. Eq.~\eqref{eq:tunneling}] for parameters $\eta=0.01$, $\Omega=10$, $\gamma=1$. $J_{QDT}(\omega)$ is shown in the inset.
Exact results for the time evolution of coherences are shown by the solid black line, and results of the different PT-MPO algorithms are indicated by points. Although an internal time step $\Delta t=1/32$ is used, data points are only shown at times $t_j=(1/4)j$ for clarity.
(b) Time evolution for $\eta=0.1$ and $\Delta t=1/128$, where the PT-MPO results are not converged. The divide-and-conquer algorithm with equal thresholds $\epsilon$ for forward and backward sweeps (shown in this case as thin green solid line) converges less regularly than the sequential algorithm; this can be ameliorated by choosing a non-unity threshold ratio $r_s=0.1$ in the preselection stage
of the divide-and-conquer block combination.
(c) Computation times of the sequential and the divide-and-conquer algorithms for different coupling strengths $\eta$ as a function of the truncation threshold $\epsilon$.
(d) Maximal inner PT dimension encountered at intermediate steps in the respective algorithm.
}
\end{figure*}

\subsubsection{Motivation and model}
To determine the practical limits of what our divide-and-conquer 
approach can achieve, we consider strong coupling to a bath with strongly peaked spectral density.
Such environments can be addressed by approximate methods such as the reaction coordinate approach~\cite{Garg1985,IlesSmith2014_environmental}, or by
specialized numerically exact techniques like 
Hierarchical Equations Of Motion (HEOM)~\cite{HEOM89}. However, such environments have been challenging for PT-MPO-based approaches due to 
long memory times and correspondingly large inner PT-MPO dimensions.
As such, they can serve as a critical test for new PT-MPO algorithms.

To explore the behavior of such models, we consider the free decay dynamics of the spin-boson model.  That is, we consider a two-level system, with vanishing system Hamiltonian,  $H_S=0$, and system-environment
coupling operator $\hat{O}=|e\rangle\langle e|$.
We then consider the decay
of initially prepared coherences, i.e.~the time evolution of
$\langle \sigma^x(t)\rangle$ for initial state
$\bar{\rho}(0)=\frac 12 (|g\rangle+|e\rangle)(\langle g|+\langle e|)$. 
This setup is particularly well suited for studying the convergence as the coherences react much more sensitively to the environment than occupations. Furthermore, for $H_S=0$ the model reduces to the independent boson model, for which analytical solutions are available for comparison. Moreover, as $[H_S,H_E]=0$, the Trotter error, i.e.~the error due to the finiteness of the time step $\Delta t$, vanishes.

We take the spectral density to have the form:
\begin{align}
\label{eq:tunneling}
J_{QDT}(\omega)=&\frac{\eta\omega\Omega^4}{(\Omega^2-\omega^2)^2+4\omega^2\gamma^2}
\end{align}
defined by interaction strength $\eta$, central frequency $\Omega$, 
and damping $\gamma$. 
This form of spectral density is commonly used to describe 
quantum dissipative tunneling~\cite{Garg1985,T-TEDOPA}, such as in 
charge or excitation transfer in biological or chemical molecular systems.
It can be derived from a hierarchical model, where the two-level system couples to a single nuclear coordinate with frequency $\Omega$ which in turn is damped by an Ohmic bath~\cite{Garg1985}.
In this interpretation,  $\eta$ is set by 
the Ohmic damping strength, and $\gamma=\eta/(2M)$ where $M$ is the nuclear mass.
Throughout this example, we set $\hbar=1$ and use dimensionless parameters 
$\Omega=10$ and $\gamma=1$, while we vary the coupling strength $\eta$.
$J_{QDT}(\omega)$ is depicted in the inset of Fig.~\ref{fig:coherence_decay}(a). 
The initial bath temperature is set to $T=0$.

\subsubsection{Numerical results and computation time}

Figure~\ref{fig:coherence_decay}(a,b) shows the time-evolution of the coherence,
calculated using the sequential and the divide-and-conquer algorithm, for $\eta=0.01, \Delta t=1/32$ 
and $\eta=0.1, \Delta t=1/128$, respectively.
For the case $H_S=0$ that we consider, exact results can be obtained 
by polaron transformation~\cite{Review_Nazir}
\begin{align}
\langle \sigma_x\rangle=\exp\bigg\{\int\limits_0^\infty d\omega
\frac{J(\omega)}{\omega^2}(\cos \omega t -1)\coth(\beta\omega/2)\bigg\}, 
\end{align}
where $\coth(\beta\omega/2)\to 1$ for zero temperature $T\to 0$.
These results are shown for comparison in Figure~\ref{fig:coherence_decay}.
For the weaker coupling, we see both methods are capable of recovering the exact result when the compression threshold $\epsilon$ is small enough.  For the stronger coupling strength convergence is not reached, as discussed further below.

Of particular interest here is how the computation time for the construction of the PT-MPOs scales as a function of
the compression threshold $\epsilon$.
This is presented in Fig.~\ref{fig:coherence_decay}(c) for baths with 
coupling strengths $\eta=0.01$, $\eta=0.04$, and $\eta=0.1$ 
and time steps $\Delta t=1/32$. 
For weak coupling $\eta=0.01$, we find the divide-and-conquer scheme
to be about one order of magnitude faster than the sequential algorithm for
all relevant compression thresholds. As shown in Fig.~\ref{fig:coherence_decay}(a) for this coupling,
both methods reproduce exact results for small thresholds $\epsilon=10^{-9}$,
while they incur comparable numerical errors for large thresholds 
$\epsilon=10^{-6}$.
 
Increasing the coupling to $\eta=0.04$, the computation time 
required for the divide-and-conquer scheme is still shorter than for
the sequential algorithm when the threshold $\epsilon$ remains large enough. 
However, the divide-and-conquer computation time increases faster with 
decreasing threshold than the time for the sequential algorithm. This increase eliminates the computational advantage over the 
sequential approach at about $\epsilon\lesssim 10^{-9}$. 
This increase in computation time corresponds to an increase in the maximal inner dimension of the PT-MPO at intermediate steps of the algorithm, as is shown in Fig.~\ref{fig:coherence_decay}(d).
From this figure we may note that the bond dimension resulting from the divide-and-conquer calculation is notably larger than for the sequential algorithm at the same cutoff.  This suggests some inefficiency in how the truncation is applied in the divide-and-conquer algorithm,  and so it may be possible to improve performance further with more sophisticated approaches to truncation~\cite{Gray2021Hyper,Banuls2023:RouteMap}.

Increasing the coupling further to $\eta=0.1$, we again see a crossover in computation times on reducing $\epsilon$, but in this case this occurs at about $\epsilon=3\times 10^{-6}$.
Once again, this increased computation time is associated with a larger maximal inner dimension.
Moreover for this coupling strength, Fig.~\ref{fig:coherence_decay}(b) shows that the convergence behavior of the divide-and-conquer 
algorithm can be less regular than that of the sequential algorithm: we see large errors at late time.
This is due to the fact that the preselection of products of singular values
does not provide the locally optimal low-rank approximation,
as discussed in detail in appendix~\ref{app:ratio}.
There, we also demonstrate that this can be combatted by using a 
finer threshold for the selection process, characterized by the ratio
$r_s={\epsilon_\textrm{select}}/{\epsilon_{\textrm{forward}}}$. Setting 
$r_s=0.1$ in simulations in Fig.~\ref{fig:coherence_decay}(b), 
we see the results of the divide-and-conquer approach match those of the sequential algorithm at $\epsilon=10^{-6}$.
We note however that neither algorithm has converged to the exact result for such a value of $\epsilon$.

Moreover we find that the choice of $r_s=0.1$ can lead to an order-of-magnitude
speedup of the divide-and-conquer algorithm for simulations with strong 
system-environment coupling $\eta=0.1$ depicted 
in Fig.~\ref{fig:coherence_decay}(c), rendering the computation time of 
the divide-and-conquer scheme close to that of the
sequential algorithm, even for the most challenging parameter regime
with small truncation thresholds.

The identification of the main drawback of our divide-and-conquer algorithm
suggests that even more challenging open quantum systems with strong 
system-environment coupling and very long memory times can be tackled by 
developing and implementing higher-quality low-rank approximations 
for the combination of PT-MPO blocks with large inner dimensions.

\section{\label{sec:summary}Summary}

We have introduced an approach for numerically exact simulations of non-Markovian open quantum systems with a scaling advantage over established techniques. 
Most techniques that account for non-Markovian effects scale quadratically $\mathcal{O}(n^2)$ with the number of times steps $n$, or linearly $\mathcal{O}(nn_c)$ if the memory can be truncated after $n_c$ time steps.
In contrast, we arrive at a method that calculates the PT-MPO---a compact representation of the Feynman-Vernon influence functional---in $\mathcal{O}(n\log n)$ or $\mathcal{O}(n_c\log n_c)$ rank reducing operations for situations without and with memory truncation, respectively. 
Once the PT-MPO is obtained, the system dynamics are simulated with linear complexity $\mathcal{O}(n)$ with a small prefactor.
We achieve this enhancement by exploiting the high degree of self-similarity in the tensor network representing the influence functional, which enables two novel developments: a divide-and-conquer strategy to contract PT-MPOs with a large number of time steps, and the construction of periodic PTs with blocks that can be repeated indefinitely. 
To demonstrate that the theoretical scaling advantage can be realized in practice, we have applied our approach to several examples of challenging multiscale problems of technological interest, and compared it with the established sequential PT-MPO calculation scheme introduced by J{\o}rgensen and Pollock~\cite{JP}.

For calculations of the fluorescence specturm of a semiconductor QD strongly coupled to a superohmic phonon bath, we indeed find a good agreement between the theoretically expected scaling behavior. We observe sublinear scaling over a wide range of practically relevant propagation times with $n=10^{3}$ to $n=10^{6}$ time steps.
The combination of divide-and-conquer with periodic PTs enables well converged calculations of the resonance fluorescence spectra of driven QDs, which requires simulations of quantum dynamics over millions of time steps. Our method produces these spectra within minutes, whereas the current standard PT algorithm of Ref.~\cite{JP} would require weeks to months to achieve the same results.

In the example of superradiant emitters, we demonstrate several additional aspects enabled by our algorithm: First, the possibility of resolving many time steps facilitates the treatment of problems in quantum optics beyond the RWA.  The PT-MPO approach allows combination of multiple environments, which facilitates investigations of the breakdown of superradiance with additional couplings to vibrational environments. 
Such numerically exact studies including effects of local environments are relevant for current efforts by several research groups to realize cooperative emission~\cite{CoopSciAdv,CygorekCoop} in spectrally tunable semiconductor QDs~\cite{Waks_superradiance,Gammon_superradiance,Lodahl_superradiance}, as well as in organic systems~\cite{SinglePhotonSuperrad}.
The ability of the divide-and-conquer scheme to explicitly model both optical decay and phonon dephasing promises tremendous acceleration of investigations of a whole class of challenging scenarios, for example, where photonic environments as well as phonon baths are strongly structured and non-additive cross-interactions between different environments are important~\cite{twobath,Nazir_nonadditive}.

While our first two examples clearly show the significant practical utility of our novel approach, a detailed analysis of performance hints at a potential disadvantage of the divide-and-conquer algorithm. It requires the combination PTs with large inner dimensions. We have shown how to significantly reduce the cost associated with this by preselection degrees of freedom based on SVDs of the individual PTs. This combination strategy however does not lead to a locally optimal rank reduction like a SVD of the combined PT. This can result in the appearance of a larger number singular values above the compression threshold than are found for the PT-MPO in the sequential algorithm of Ref.~\cite{JP}. Consequently, the inner dimension $\chi$ of the PT may be increased, and thereby the computation time needed to perform each SVD.
This limitation of the divide-and-conquer approach is illustrated in our final example:  free coherence decay with a model spectral density consisting of a narrow peak. This choice is motivated by the expectation of large inner dimensions $\chi$, which are controlled by the bath coupling strength in such a case. Indeed, we find that on increasing the coupling strength, there is a cross-over from a regime where the divide-and-conquer algorithm outperforms the sequential algorithm to a regime where the sequential algorithm performs better. Simple strategies to mitigate the extra singular values appearing in the divide-and-conquer scheme, e.g., by using different compression thresholds at different steps in the algorithm, are promising. 

Such numerical experiments support a picture that the optimal strategy for PT-MPO calculations is a balance between reducing the number of SVDs (or alternative rank-reducing operations) and reducing the numerical effort of each individual operation. 
The divide-and-conquer algorithms provides a scaling advantage for the number of operations, but at the cost of increasing the effort of each operation.
It should however be noted that the increased inner PT dimension of the divide-and-conquer algorithm is not due to the divide-and-conquer strategy \textit{per se}:  
it is a consequence of the suboptimal compression after joining two PT-MPOs with large inner dimensions.
It would thus be worthwhile for future work to explore other strategies to combine large PT-MPOs, e.g., alternate criteria for selecting singular values, or other methods for low-rank matrix approximations~\cite{Gray2021Hyper,Banuls2023:RouteMap}.
Alternatively, one may also construct approaches that interpolate between the sequential and the divide-and-conquer approach. 
For example,  merging fixed-size blocks of the tensor network instead of single lines in the sequential algorithm, or by sweeping and compressing PT-MPOs multiple times after combining into blocks, and allowing a different compression threshold for each sweep. 
We also note that periodic PTs can be readily calculated starting from blocks obtained with the sequential algorithm. 
This could lead to sizeable reduction of computation time when propagating open quantum systems coupled to environments with finite memory times but very strong system-environment couplings.

Finally, it is noteworthy that, even though we have implemented the divide-and-conquer approach only for bosonic environments, as defined in Eq.~\eqref{eq:spinboson}, in principle it can be extended to the more general class of Gaussian environments~\cite{twobath,StrathearnBook}, which includes, e.g., fermionic environments of impurity problems~\cite{PT_Abanin,PT_Reichman}. 
Moreover, PTs describing environments coupled to small quantum systems can be reused for simulations for which the system of interest is coupled to another subsystem. 
Extending the system Hilbert space is accommodated simply by modifying the outer bonds of the PT-MPO~\cite{ACE,PI_cavityfeeding}. Thus, the PT-MPO calculated in our first example can be readily employed to obtain spectra of QDs embedded in microcavities, which is another current topic of interest~\cite{Hughes_QDcavitySpectra}.
A variant of this approach was used in Ref.~\cite{CoopWiercinski} to investigate cooperative emission from two QDs, each additionally strongly coupled to a local phonon environment, providing numerically exact predictions including possible cross-interactions between different environments. Furthermore, Ref.~\cite{Fux_spinchain} delivers a proof of principle for scaling up PT-MPO-based numerically exact approaches to small networks of open quantum systems. 
Given this wider context, progress in methods for constructing PT-MPOs has immediate implications for a large class of topical problems in open quantum systems.

After submission of our manuscript, a related preprint~\cite{Strunz_infinite} by Link \emph{et al.} showed how periodic PTs can be also be calculated using iTEBD methods. 

\acknowledgements
M.C. and E.M.G. acknowledge funding from EPSRC grant no. EP/T01377X/1. 
B.W.L. and J.K. were supported by EPSRC grant no. EP/T014032/1. We also thank Gerald Fux for fruitful discussions.

\appendix
\section{Different thresholds for selection, backward, and forward sweeps\label{app:ratio}}

We now discuss how choosing different thresholds for
forward and backward sweeps as well as for the selection of products of 
singular values can affect the performance of the divide-and-conquer algorithm.
To this end, we first review the role of canonical forms for 
the optimality of matrix product state (MPS) 
compression~\cite{MPS_Schollwoeck, Cirac_faithful}.
We then discuss the combination based on selecting products of 
singular values, before providing numerical examples.

\subsection{Locally optimal MPS truncation}
A general pure state $|\psi\rangle$ of a one-dimensional quantum system
with $L$ sites, each described by a Hilbert space of dimension $d$, can 
be expanded as
\begin{align}
|\psi\rangle = \sum_{\sigma_1,\dots,\sigma_L} c_{\sigma_1,\dots,\sigma_L}
|\sigma_1,\dots,\sigma_L\rangle,
\end{align}
where $|\sigma_1,\dots,\sigma_L\rangle$ are products of local basis states
and $c_{\sigma_1,\dots,\sigma_L}$ are the $d^{L}$ expansion coefficients.
These can be exactly expressed as matrix products
\begin{align}
\label{eq:MPS}
c_{\sigma_1,\dots,\sigma_L}=\sum_{a_1=1}^{\chi_1} \dots
\sum_{a_{L-1}=1}^{\chi_{L-1}} M^{\sigma_1}_{a_0,a_1}
M^{\sigma_2}_{a_1,a_2} \dots M^{\sigma_L}_{a_{L-1},a_L},
\end{align}
where $a_0=a_L=1$ are dummy indices. The dimensions $\chi_L$ of 
indices $a_l$ in general grow exponentially towards the center, e.g.,
$\chi_{L/2}\le d^{L/2}$ for even $L$.
To approximate the state $|\psi\rangle$ by an MPS 
$|\psi_\textrm{trunc}\rangle$ with truncated bond dimension $D$ between 
sites $l$ and $l+1$,
it is advisable to first use the gauge freedom of the inner bonds to bring the
MPS into the mixed-canonical form~\cite{MPS_Schollwoeck}
\begin{align}
&c_{\sigma_1,\dots,\sigma_L}\nonumber\\
&=\sum_{a_1,\dots,a_{L-1}}
A^{\sigma_1}_{a_0,a_1}A^{\sigma_2}_{a_1,a_2} \dots A^{\sigma_l}_{a_{l-1},a_l}
{s}_{a_l}
B^{\sigma_{l+1}}_{a_l, a_{l+1}} B^{\sigma_L}_{a_{L-1},a_L}
\label{eq:mixed_canonical}
\end{align}
with 
$\sum_{\sigma}A^{\sigma \dagger} A^{\sigma} = \mathbb{1}$ and
$\sum_{\sigma}B^{\sigma} B^{\sigma\dagger} = \mathbb{1}$.
With these conditions, the Schmidt decomposition at the bond between sites 
$l$ and $l+1$ is given by
\begin{align}
|\psi\rangle=\sum_{a_l=1}^{\chi_l} s_{a_l} |a_l\rangle_A |a_l\rangle_B 
\label{eq:schmidt}
\end{align}
with bases for the left and right blocks
\begin{align}
|a_l\rangle_A =& \sum_{\sigma_1,\dots,\sigma_l} 
(A^{\sigma_1}\dots A^{\sigma_l})_{1,a_l} 
|\sigma_1,\dots,\sigma_l\rangle,\\
|a_l\rangle_B =& \sum_{\sigma_{l+1},\dots,\sigma_L} 
(B^{\sigma_{l+1}}\dots B^{\sigma_L})_{a_l,1} 
|\sigma_{l+1},\dots,\sigma_L\rangle,
\end{align}
respectively, where the orthonormality
${}_A\langle a_l| a'_l\rangle_A=
{}_B\langle a_l| a'_l\rangle_B=\delta_{a_l, a'_l}$
of the Schmidt basis follows directly from the orthonormality of
$A^{\sigma_l}$ and $B^{\sigma_l}$.
Given the Schmidt decomposition~\eqref{eq:schmidt},
the locally optimal approximation to $|\psi\rangle$ by a MPS with reduced 
dimension $D$, guaranteed by the Eckart-Young-Mirsky theorem~\cite{eckart1936},
is obtained by neglecting the smallest singular values
\begin{align}
|\psi_\textrm{trunc}\rangle=\sum_{a_l=1}^{D}s_{a_l}|a_l\rangle_A|a_l\rangle_B,
\end{align}
which is equivalent to restricting the sum over $a_l$ in the mixed-canonical
MPS representation in Eq.~\eqref{eq:mixed_canonical} to the first $D$ terms.
The truncation error at a single site is
\begin{align}
E_l(D)=\| |\psi\rangle-|\psi_\textrm{trunc}\rangle \|_2^2 = 
\sum_{a_l=D+1}^{\chi_l} s_{a_l}^2.
\end{align}
A general MPS as in Eq.~\eqref{eq:MPS} can be brought into
the mixed-canonical form of Eq.~\eqref{eq:mixed_canonical} by sweeping, e.g.,
in the forward direction (from $l=1$ to $l=L-1$) while performing SVDs.
There the MPS matrices are replaced by
\begin{subequations}
\begin{align}
M^{\sigma_l}_{a_{l-1},a_l} &\longrightarrow 
U_{(\sigma_l,a_{l-1}),\tilde{a}_l}s_{\tilde{a}_l}V^\dagger_{\tilde{a}_l,a_l},\\
A^{\sigma_l}_{a_{l-1},\tilde{a}_l}\equiv
M^{\sigma_l}_{a_{l-1},\tilde{a}_l} &\longleftarrow 
U_{(\sigma_l,a_{l-1}),\tilde{a}_l},\\
M^{\sigma_{l+1}}_{\tilde{a}_{l},{a}_{l+1}} &\longleftarrow 
\sum_{a_l} s_{\tilde{a}_l}V^\dagger_{\tilde{a}_l,a_l}
M^{\sigma_{l+1}}_{{a}_{l},{a}_{l+1}}.
\end{align}
\end{subequations}
The normalization property 
$\sum_{\sigma}A^{\sigma \dagger} A^{\sigma} = \mathbb{1}$ 
for MPS matrices left of the link $l$ during the forward sweep 
follows from the orthonormality of the 
columns of $U_{(\sigma_l,a_{l-1}),\tilde{a}_l}$. 
Analogously, a backward sweep (from $l=L$ to $l=2$)
creates the orthonormality condition 
$\sum_{\sigma}B^{\sigma} B^{\sigma \dagger} = \mathbb{1}$
for all MPS matrices right of the link $l$ during the sweep.

To summarize, the locally optimal compression of a non-canonical MPS
in principle requires
a first sweep along one direction without truncation to restore
the canonical form, before the inner bond dimensions are reduced in a 
subsequent sweep along the opposite direction with truncated SVDs.
The overall error is then bounded by~\cite{Cirac_faithful}
$\| |\psi\rangle-|\psi_\textrm{trunc}\rangle \|_2^2 \le 2\sum_{l=1}^L E_l(D)$.
These results for MPS similarly apply to MPOs with only subtle differences, 
such as that MPOs are generally not normalized to unity and may require 
rescaling~\cite{Hubig_MPO}.

\subsection{Backward versus forward sweep thresholds}
Despite the fact that the PT-MPO is not of canonical form 
after incorporating another row of the tensor network into the PT-MPO, 
it is standard practice for the sequential algorithm~\cite{JP}
to truncate small singular values in every sweep.
Empirically, this is justified by the observation that the
potentially suboptimal low-rank approximation---and thus 
larger bond dimension of PT-MPOs after compression for the same accuracy---is 
typically overcompensated by the reduced computation time due to much smaller bond dimensions during the second sweep. 

Here, we investigate whether a similar practice is also advisable 
for the divide-and-conquer algorithm. 
There, the combination of blocks takes place before the backward sweep,
i.e.~the backward sweep is responsible for restoring the canonical form.
By choosing different compression thresholds for backward and forward sweeps, 
which we characterize by the ratio 
${r}_b=\epsilon_\textrm{backward}/\epsilon_\textrm{forward}$, we
interpolate between the choices of equal thresholds $r_b=1$, as in the common
practice for the sequential algorithm~\cite{JP}, 
and $r_b\to 0$, where the MPO is only brought to canonical form
so that the truncation during the subsequent forward sweep is locally optimal.
For fixed forward sweep threshold $\epsilon_\textrm{forward}=\epsilon$, the
latter can lead to smaller inner bond dimensions compared to calculations with 
$r_b=1$ and, hence, to an overall faster algorithm. 
The optimal choice for $r_b$ in different scenarios are explored below in 
numerical experiments.

\subsection{Preselection thresholds}
The preselection process in Eq.~\eqref{eq:select_product} consists of combining MPO matrices 
$\big[\tilde{f}^{\alpha_{l}}_{k:j}\big]_{d'_l,d'_{l-1}}$ and
$\big[\tilde{g}^{\alpha_l}\big]_{d''_l,d''_{l-1}}$
by first constructing SVDs of the individual matrices
$\big[\tilde{f}^{\alpha_{l}}_{k:j}\big]_{d'_l,d'_{l-1}}=
\sum_s U^{(1)}_{d'_l, s} \sigma^{(1)}_s V^{(1)\dagger}_{s, (\alpha_l,d'_{l-1})}$
and
$\big[\tilde{g}^{\alpha_l}\big]_{d''_l,d''_{l-1}}=
\sum_t U^{(2)}_{d''_l, t} \sigma^{(2)}_t V^{(2)\dagger}_{t, (\alpha_l,d''_{l-1})}$.
Then, combined indices $(s,t)$ corresponding to 
small products of singular values 
$\sigma_s^{(1)}\sigma_t^{(2)}< \epsilon_\textrm{select}\sigma_0^{(1)}\sigma_0^{(2)}$ are disregarded. This  process also does not result
in the locally optimal low-rank approximation for a given bond dimension.
However, it is noteworthy that if the outer bonds $\alpha_l$ on the combined
MPO matrices were to act on different spaces described by different indices 
$\alpha_l$ and $\alpha'_l$, respectively, the decomposition of the matrix
\begin{align}
&M_{(d'_l,d''_l),((\alpha_l,d'_{l-1}),(\alpha'_l,d''_{l-1}))}=
\big[\tilde{f}^{\alpha_{l}}_{k:j}\big]_{d'_l,d'_{l-1}}
\big[\tilde{g}^{\alpha'_l}\big]_{d''_l,d''_{l-1}}
\nonumber\\&=
\sum_{s,t}
\big(U^{(1)}_{d'_l, s} U^{(2)}_{d''_l, t} \big)
\sigma^{(1)}_s \sigma^{(2)}_t
\big(V^{(1)\dagger}_{s, (\alpha_l,d'_{l-1})}
V^{(2)\dagger}_{t, (\alpha'_l,d''_{l-1})}\big)
\label{eq:select_aaprime}
\end{align}
is---up to reordering of singular values---the SVD $M=U \Sigma V^\dagger$ with 
$U=U^{(1)}\otimes U^{(2)}$, $V^\dagger=V^{(1)\dagger} \otimes V^{(2)\dagger}$
and $\Sigma=\Sigma^{(1)}\otimes \Sigma^{(2)}$
with $\Sigma^{(i)}=\textrm{diag}(\sigma^{(i)})$ the diagonal matrices 
containing the individual singular values. 
Hence, by the Eckart-Young-Mirsky theorem~\cite{eckart1936}, our truncation
procedure based on products of singular values is optimal in this expanded 
space.
The combined matrix 
$\big[\tilde{f}^{\alpha_{l}}_{k:j+1}\big]_{(d'_l,d''_l),(d'_{l-1},d''_{l-1})}=
\big[\tilde{f}^{\alpha_{l}}_{k:j}\big]_{d'_l,d'_{l-1}}
\big[\tilde{g}^{\alpha_l}\big]_{d''_l,d''_{l-1}}$ 
in Eq.~\eqref{eq:select_product} can be obtained by eliminating one outer bond
\begin{align}
&\big[\tilde{f}^{\alpha_{l}}_{k:j+1}\big]_{(d'_l,d''_l),(d'_{l-1},d''_{l-1})}
\nonumber\\
&=\sum_{\alpha'_l}\delta_{\alpha_l,\alpha'_l}
M_{(d'_l,d''_l),((\alpha_l,d'_{l-1}),(\alpha'_l,d''_{l-1}))}.
\label{eq:select_aadelta}
\end{align}
Hence, what makes our selection procedure suboptimal is that it neglects the 
information that only the elements of $M$ with $\alpha'_l=\alpha_l$ have to
be reproduced by the low-rank approximation. Again, one strategy to 
compensate for this is to choose a smaller threshold $\epsilon_\textrm{select}$ 
for the selection of products of singular values,
which we characterize by the ratio 
${r}_s=\epsilon_\textrm{select}/\epsilon_\textrm{forward}$.

\subsection{Numerical study of optimal threshold ratios}
\begin{figure}
\includegraphics[width=0.99\linewidth]{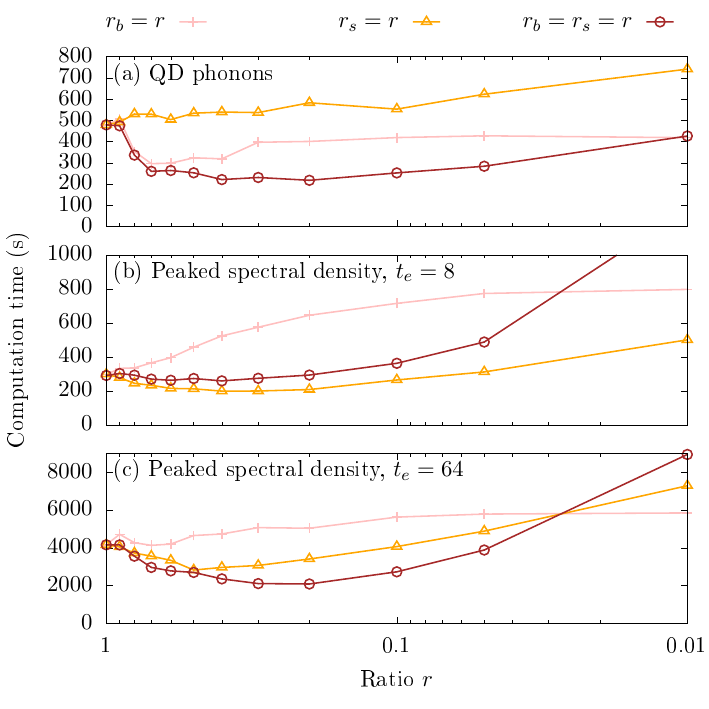}
\caption{Computation time for PT calculation 
as a function of ratios between backward versus forward sweeps threshold
$r=r_b=\epsilon_\textrm{backward}/\epsilon_\textrm{forward}$,
selection threshold versus forward sweep threshold
$r=r_s=\epsilon_\textrm{select}/\epsilon_\textrm{forward}$,
and both, $r=r_b=r_s$.
Results shown for (a) a quantum dot 
coupled to phonons as in section~\ref{sec:QDPhonon}
and strongly peaked spectral densities as in section~\ref{sec:coherence_decay}  
with coupling strength $\eta=0.01$ 
for total propagation times $t_e=8$ (b) and $t_e=64$ (c), respectively. 
\label{fig:ratios}}
\end{figure}

We now investigate how the overall computation time of PT-MPO 
simulations is affected by the threshold ratios $r_b$ and $r_s$.
Fig.~\ref{fig:ratios}a shows results for PT-MPOs for a quantum dot
coupled to longitudinal acoustic phonons as in section~\ref{sec:QDPhonon}.
The PT-MPO is calculated by the divide-and-conquer algorithm using the 
same spectral density as in Fig.~\ref{fig:QDPhonon},
threshold $\epsilon_\textrm{forward}=\epsilon=10^{-12}$, 
time step $\Delta t=0.01$ ps,
and total propagation time $t_e=20.48$~ps.

First of all, we find that reducing only the backward sweep threshold
ratio $r_b$ already results in faster 
PT-MPO construction with optimal ratio around $r_b\sim 0.7$. 
The computation time can be further reduced by simultaneously decreasing the
threshold used for preselection via $r_s$ 
where less than half the computation time is required
for the choice $r_s=r_b=0.2$ compared to equal thresholds $r_s=r_b=1$.
In contrast, here, decreasing only $r_s$ does not lead to a better overall 
performance. This suggests that, in the example studied here, the 
preselection is already nearly optimal in selecting relevant degrees of freedom,
while the MPO after selection and combination deviates considerably 
from its canonical form.
The former can be explained by the fact that what dominates the 
computation time are the last few iterations of the divide-and-conquer 
algorithm, where the inner dimensions are largest.
In this example, this involves the combination of elements that are 
shifted by a number of time steps of the order of the memory time $n_c$,
where the new block overlaps with the long tail of the MPO with elements that
are nearly independent of the outer index $\alpha_l$,
as $[b_{l}^{\alpha_i,\alpha_j}]\to 1$ for $l\to n_c$, irrespective of 
the value of $\alpha_i$ and $\alpha_j$. 
In this case, it can be expected that the difference between
Eq.~\eqref{eq:select_aaprime} and Eq.~\eqref{eq:select_aadelta} 
becomes less relevant.
Moreover, we attribute the observation that decreasing $r_s$ along with 
$r_b$ is more efficient than the decrease of $r_b$ alone to consistency:
When $\epsilon_\textrm{select}<\epsilon_\textrm{backward}$, terms are 
eliminated in a not locally optimal way during the preselection that would
otherwise allow the canonicalization during the backward sweep to result 
in a more compact and accurate form. 

A somewhat different picture unfolds for the example of the spin-boson 
model with strongly peaked spectral density with coupling strength 
$\eta=0.01$ and total propagation time $t_e=8$ discussed in 
section~\ref{sec:coherence_decay}, for which we plot the computation time
for varying threshold ratios $r_b$ and $r_s$ in
Fig.~\ref{fig:ratios}b.
There, decreasing only $r_b$ increases the computation time, while 
decreasing $r_s$ is found to be more efficient. 
The main difference to the situation in Fig.~\ref{fig:ratios}a
is that in the parameter regime considered in Fig.~\ref{fig:ratios}b
the final time $t_e$ is still well within the memory time of the environment,
and the preselection combines MPO matrices whose elements depend significantly
on the outer indices. As such, optimal compression of
Eq.~\eqref{eq:select_aaprime} does not necessarly compress 
Eq.~\eqref{eq:select_aadelta} optimally, and 
the preselection based on products of singular
values is not a perfect proxy for the SVD of the product of the 
MPO matrices.
As small $r_s$ increases the inner bond dimension after the selection process, 
it appears that truncating sooner rather than later, i.e.~already during
the backward sweep by keeping $r_b=1$, is the best strategy to minimize the
overall computation time in this situation.

To test the hypothesis that the truncation strategy should be chosen differently
depending on whether the total propagation time exceeds the physical
memory time of the environment, we show in Fig.~\ref{fig:ratios}c the
computation times for simulations as in Fig.~\ref{fig:ratios}b but with
larger total propagation time $t_e=64$. In this case, we find indeed that
the optimal strategy is to reduce both threshold ratios $r_s$ and $r_b$, as
the situation is more comparable to the one in Fig.~\ref{fig:ratios}a.

\section{\label{app:spectra_convergence}Convergence of fluorescence spectra}
\begin{figure}
\includegraphics[width=0.99\linewidth]{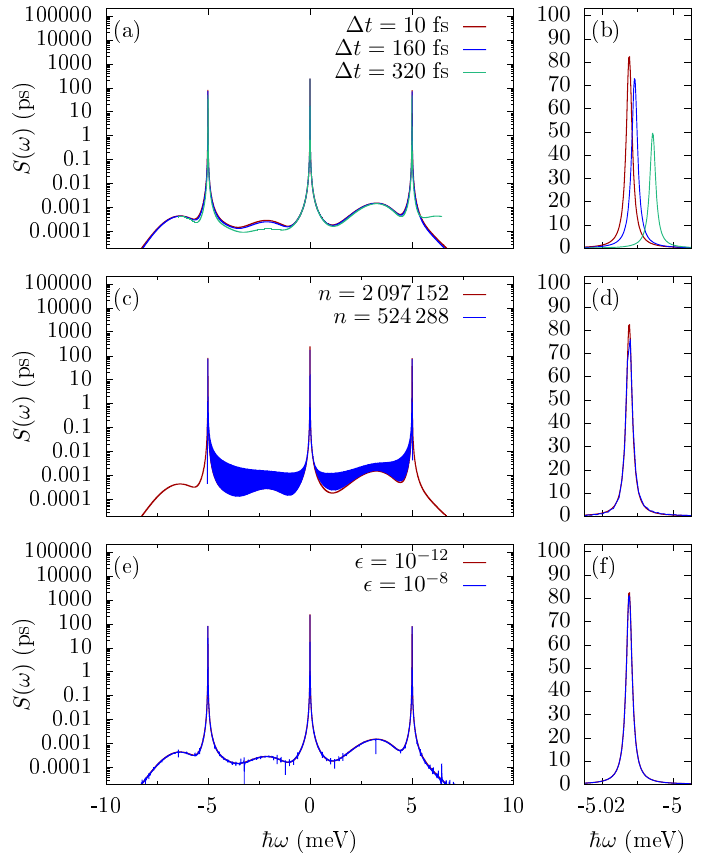}
\caption{Convergence of fluorescence spectra with respect to time step 
$\Delta t$ (a,b), total number of time steps $n$ (c,d), and truncation threshold
$\epsilon$ (e,f). Left panels (a,c,e) use logarithmic scale, right panels 
(b,d,f) show a zoom into the left peak on a linear scale. 
The remaining parameters are the same as for Fig.~\ref{fig:QDPhonon}e.
\label{fig:spectra_convergence}}
\end{figure}

To demonstrate that calculating fluorescence spectra of quantum dots in
the strong driving limit indeed puts strong requirements on the convergence
parameters, we compare in Fig.~\ref{fig:spectra_convergence} the
spectra obtained for $\hbar\Omega=5$ meV in Fig.~\ref{fig:QDPhonon}e with
simulations where a single convergence parameter is varied. 

The impact of a too coarse time discretization can be observed in 
Fig.~\ref{fig:spectra_convergence}a and b, where panel a shows the spectra
on logarithmic scale while panel b shows the left Mollow peak enlarged and
on a linear scale. The overall memory time $n_c \Delta t=20.48$ ps is kept fixed, i.e.~$n_c$ is varied along with the time steps $\Delta t$. For larger time steps $\Delta t=320$ fs, one already finds many features of the spectra well reproduced on the logarithmic scale. Note, however, that sizable deviations are still found on the linear scale in panel b for $\Delta t=160$, suggesting that our
choice for $\Delta t=10$ fs in the main text is indeed of the required order 
of magnitude for convergence.
In additional calculations (not shown), we kept this time step $\Delta t=10$~fs while varying the memory cut-off $n_c$. We found the algorithm to lead to fastest results for $n_c=2\,048$ memory time steps, i.e., no gain is made by truncating the memory earlier. 

In Fig.~\ref{fig:spectra_convergence}c and d, we show that reducing the
overall number of time steps from $n=2^{21}$ to $n=2^{19}$ 
results in oscillatory artefacts in the spectra, as the total propagation
time $t_e=n\Delta t$ is too short for all itinerant excitations in the two-time
correlation function to decay.
One can also see on the linear scale in panel d 
that the low-energy peak is not sampled sufficiently.

Finally, the impact of the compression threshold is analyzed in 
Fig.~\ref{fig:spectra_convergence}e and f.  This seems less severe, as it mainly leads
to small blips on top of the otherwise well reproduced spectra. 
Note, however, that we find deviations of the two-time correlation functions 
in the time domain by $\lesssim 10^{-4}$ when comparing simulations with
$\epsilon=10^{-12}$ to those with $\epsilon=10^{-11}$ (not shown).
This indicates that such small thresholds are required for well 
converged dynamics, yet small errors in temporal data may be averaged out 
by the Fourier transform.

In any case, the chosen convergence parameters $\Delta t=10$ fs, $n_c=2^{11}$, 
and $n=2^{21}$ used for the simulations in Fig.~\ref{fig:QDPhonon}
are indeed not too far from the minimal requirement for well converged spectra.

\section{\label{app:revrwa}Reverse rotating wave approximation}

In this appendix we discuss how one may consider Eq.~\eqref{eq:Hsup_orig} as a ``reverse rotating wave approximation'' of the standard model of superradiance, that uses the Tavis--Cummings model.  In this picture we consider using the reverse RWA to convert a system-environment coupling that is not of the product form required to match Eq.~\eqref{eq:spinboson} into such a form.  In such a picture, we consider $\omega_R$ as a convergence parameter that determines the quality of the approximation.

We start by summarising the standard RWA, starting from Eq.~\eqref{eq:Hsup_orig}.
Such an approximation is valid if the typical value of the emitter frequencies, $\omega_j$, is much larger than
the couplings $g_k$. The RWA is achieved by first applying a time-dependent 
unitary to change to a rotating frame,
\begin{align}
U(t) = e^{i \omega_{R} (\sum_j \sigma^+_j \sigma^-_j + 
\sum_k a^\dagger_k a_k ) t }.
\end{align}
Here $\omega_R$ is a reference frequency, typically chosen to be similar to
the emitter frequencies $\omega_j$.
If a wave function $|\Psi(t)\rangle$ 
obeys the Schr{\"o}dinger equation 
$i\hbar \frac{\partial}{\partial t}|\Psi(t)\rangle =H |\Psi(t)\rangle$,
then \mbox{$|\Psi'(t)\rangle= U(t) |\Psi(t)\rangle$}  obeys
\begin{align}
i\hbar \frac{\partial}{\partial t}|\Psi'(t)\rangle =
\Big(U H U^\dagger +i\hbar \frac{\partial U}{\partial t} U^\dagger \Big)
|\Psi'(t)\rangle = {H}' |\Psi'(t)\rangle
\end{align}
with
\begin{align}
\label{eq:Hsup_afterTrafo}
H'=& \sum_{j=1}^N \hbar (\omega_j-\omega_R) \sigma^+_j \sigma^-_j 
+\sum_k \hbar (\omega_k-\omega_R) a^\dagger_k a_k 
\nonumber\\&
+\sum_{j,k} \hbar g_k\big( a^\dagger_k \sigma^-_j+ a_k \sigma^+_j \big) 
\nonumber\\&
+\sum_{j,k} \hbar g_k\big( a^\dagger_k \sigma^+_j e^{i2\omega_R t}
+a_k \sigma^-_j e^{-i2\omega_R t} \big).
\end{align}

The RWA is completed by neglecting the counter-rotating terms, i.e.~the
terms oscillating with frequency $\pm2\omega_R$, which is justified
if the reference frequency $\omega_R$ is much larger than any
other frequency in the system.
The impact of these terms on the dynamics on a coarse-grained 
time scale $T$ is of the order
\begin{align}
\label{eq:RWA_error_scaling}
\frac 1T\int\limits_0^{T} dt\, e^{\pm i2\omega_R t}=
\frac{e^{\pm i2\omega_R T}-1}{\pm i2\omega_RT} =
\mathcal{O}\bigg(\frac{1}{\omega_R T}\bigg).
\end{align}

With the definition of relative frequencies 
$\tilde{\omega}_j= \omega_j-\omega_R$ and 
$\tilde{\omega}_k =\omega_k-\omega_R$, the RWA turns the multi-mode Dicke model
of Eq.~\eqref{eq:Hsup_orig} into the multi-mode Tavis-Cummings model
\begin{align}
\label{eq:Hsup_RWA}
\tilde{H}=&\sum_{j=1}^N \hbar \tilde{\omega}_j \sigma^+_j \sigma^-_j 
+\sum_k \hbar \tilde{\omega}_k a^\dagger_k a_k 
\nonumber\\&
+\sum_{j,k} \hbar g_k\big(a^\dagger_k \sigma^-_j+a_k \sigma^+_j \big).
\end{align}
Eq.~\eqref{eq:Hsup_RWA} has the advantage that only comparatively slow oscillations with frequencies of the order $\tilde{\omega}_j$ and $g_k$ have to be resolved;  fast oscillations with frequencies $\omega_R$ are eliminated.  One may also note that after this process, the value $\omega_R$ does not appear:  as long as the RWA is valid,  models with different values of $\omega_R$ will all map to the same approximate model.

We may now consider this process in reverse.
If one requires a solution of the multi-mode Tavis-Cummings model in Eq.~\eqref{eq:Hsup_RWA}, a reverse rotating wave approximation can be applied by introducing a reference frequency $\omega_R$,
which now takes to role of an additional convergence parameter.
As seen in Eq.~\eqref{eq:RWA_error_scaling}, the error introduced by the reverse RWA
can be systematically reduced by increasing the frequency $\omega_R$. 
Our divide-and-conquer algorithm is key to dealing with the small time steps 
$\Delta t$ needed to resolve the fast oscillations
$\omega_R\Delta t\ll 1$.
As noted above, the value of $\omega_R$ does not enter the calculation as long as it is large enough.  This means that one may also use the above arguments to model systems where the physical $\omega_R$ is too large even for the divide-and-conquer approach.  A RWA can be made to eliminate the original common frequency, and then the reverse RWA made to recover a model of the form of Eq.~\ref{eq:spinboson}, allowing the use of the PT-MPO approaches described in this paper.

\section{\label{app:superrad_explicit}Analytic expression for photon intensities from ideal superradiant emitters}
Here, we derive analytic expressions for the intensity of photons emitted 
from $N$ ideal superradiant emitters coupled to a flat (Markovian) bath of photon modes. As described in the main text, the
dynamics is governed by a sequence of photon emission processes between
the Dicke states $|J,M\rangle$ and $|J,M-1\rangle$ with rates
$\Gamma_{J,M}=\kappa(J+M)(J-M+1)$, where $J=N/2$ and 
$M=J, J-1, \dots, -J$.
We denote by $p^{[N]}_M$ the populations of the Dicke state $|N/2, M\rangle$.
The corresponding rate equation are 
\begin{subequations}
\begin{align}
\label{eq:app_sup_pmax}
\frac{\partial}{\partial t}p^{[N]}_{N/2} =& -\Gamma_{N/2,N/2}\, p^{[N]}_{N/2}
\end{align}
for the Dicke state with maximal excitation $M=N/2$ and 
\begin{align}
\label{eq:app_sup_pother}
\frac{\partial}{\partial t}p_{M}^{[N]} =&  
\Gamma_{N/2, M+1}\, p_{M}^{[N]} - \Gamma_{N/2, M}\, p_M^{[N]} 
\end{align}
\end{subequations}
for the remaining Dicke states with $M<N/2$.
These equations are solved by 
\begin{subequations}
\begin{align}
\label{eq:app_sup_sol_max}
p^{[N]}_{N/2}(t)=&e^{-\Gamma_{N/2,N/2} t} 
\end{align}
and
\begin{align}
\label{eq:app_sup_sol_other}
p^{[N]}_{M}(t) =&
\Gamma_{N/2,M+1}\int\limits_{0}^{t}d\tau e^{-\Gamma_{N/2,M} (t-\tau)}
p^{[N]}_{M+1}(\tau),
\end{align}
\end{subequations}
respectively, where we have already incorporated the initial condition
of initially maximally excited emitters
$p^{[N]}_{N/2}(0)=1, p^{[N]}_{M<N/2}(0)=0$.
It can be seen that the solution $p^{[N]}_{M}(t)$ can be calculated by 
integrating over the solutions $p^{[N]}_{M+1}(t)$ after
multiplying with a simple exponential kernel. Hence, all Dicke state 
populations can be obtained one at a time by Eq.~\eqref{eq:app_sup_sol_other}
starting from $p^{[N]}_{N/2}(t)$ given in Eq.~\eqref{eq:app_sup_sol_max}.

Because the Dicke state $|J,M\rangle$ carries $M+J$ excitations,
the emitted photon intensity, i.e. the negative change of the
total emitter excitation, is
\begin{align}
I^{[N]}=-\frac{\partial}{\partial t} 
\sum\limits_{M=-N/2}^{N/2} (M+N/2)\, p_M^{[N]}.
\end{align}
For up to $N=5$, we thus find
\begin{subequations}
\begin{align}
\frac{I^{[1]}}{\kappa}=&
e^{-\kappa t},
\\
\frac{I^{[2]}}{2\kappa}=&
\left(2\kappa t +1\right) e^{-2\kappa t},
\\
\frac{I^{[3]}}{3\kappa}=&
\left(12\kappa t -7 \right)e^{-3\kappa t} +8e^{-4\kappa t},
\\
\frac{I^{[4]}}{4\kappa}=&
\left(36\kappa t-23\right)e^{-4\kappa t}+
\left(18\kappa t+24\right)e^{-6\kappa t},
\\
\frac{I^{[5]}}{5\kappa}=&
\left(80\kappa t-\frac{143}{3}\right)e^{-5\kappa t} + 
\left(128\kappa t-\frac{16}{3}\right)e^{-8\kappa t} 
\nonumber\\&
+54 e^{-9\kappa t}.
\end{align}
\end{subequations}

\input{paper_revised.bbl}

\end{document}

%% file: paper_revised.bbl
%